\documentclass[structabstract]{aa}  
\usepackage{color}
\usepackage{natbib,longtable}
\usepackage{graphicx}
\usepackage{txfonts}
\usepackage{epstopdf}
\bibpunct{(}{)}{;}{a}{}{,}

\newcommand{\dataset}{data set}
\newcommand{\Dataset}{Data set}
\newcommand{{\iraf}}{\texttt{IRAF}}
\newcommand{\snid}[1]{{ID #1}}

\newcommand{\nogalspectext}{Fourteen}
\newcommand{\nozsurveytext}{two}
\newcommand{\nochangetext}{Three}
\newcommand{\noremsubtext}{two}
\newcommand{\noonesnidagetext}{one}   

\newcommand{\nttnights}{32}  
\newcommand{\notnights}{9}

\newcommand{\nospectra}{290}
\newcommand{\noobj}{238}

\newcommand{\nosnspectra}{207}
\newcommand{\nosne}{172}
\newcommand{\nosniaspectra}{169}
\newcommand{\nosneia}{141}
\newcommand{\nosniaspectralc}{127}
\newcommand{\nosneialc}{108}

\newcommand{\nosneiabadlc}{33}

\newcommand{\noobsnot}{46}
\newcommand{\noobsntt}{244}
\newcommand{\nosnIIspectra}{26}
\newcommand{\nosneII}{23}
\newcommand{\nosnIbcspectra}{12}
\newcommand{\nosneIbc}{8}
\newcommand{\nogalspec}{14}
\newcommand{\nogal}{12}
\newcommand{\nonotsnspec}{19}
\newcommand{\nounknownspec}{61}
\newcommand{\nounknown}{47}
\newcommand{\nonotsn}{16}
\newcommand{\nosniaspectraq}{3}
\newcommand{\nosneiaq}{3}
\newcommand{\snproc}{72}

\newcommand{\zmean}{$0.16$}
\newcommand{\nozhist}{228}
\newcommand{\zBins}{0.03}
\newcommand{\zminall}{$0.016$}
\newcommand{\zmaxall}{$0.487$}
\newcommand{\zminia}{$0.031$}
\newcommand{\zmaxia}{$0.324$}
\newcommand{\zminiacrop}{$0.03$}
\newcommand{\zmaxiacrop}{$0.32$}
\newcommand{\noobjwz}{228}   
\newcommand{\nospecwz}{280}  
\newcommand{\noobjnoz}{10}

\newcommand{\nozsdss}{84}
\newcommand{\nozgal}{102}
\newcommand{\nozsn}{40}

\newcommand{\surveyz}{16391 and 16838} 

\newcommand{\noshist}{108}
\newcommand{\sMean}{$0.95$}

\newcommand{\sStd}{$0.11$}
\newcommand{\sBins}{$0.05$}

\newcommand{\deltamMean}{$1.13$}

\newcommand{\deltamStd}{$0.22$}

\newcommand{\nochist}{108}
\newcommand{\cMean}{$0.08$}
\newcommand{\cMedian}{$0.07$}
\newcommand{\cStd}{$0.14$}
\newcommand{\cBins}{$0.05$}

\newcommand{\noephist}{$127$}
\newcommand{\epochBins}{$3$}

\newcommand{\agedisp}{$4$} 
\newcommand{\zdisp}{$0.005$} 

\newcommand{\noonesnidageid}{20142}

\newcommand{\nohandtyped}{$36$}
\newcommand{\nofromunknown}{$20$}
\newcommand{\notounknown}{$11$}

\newcommand{\perlowcont}{31}
\newcommand{\prcontlim}{20}    

\newcommand{\medianairmassobs}{1.4}       
\newcommand{\perchighairmassobs}{39}
\newcommand{\perchighangleobs}{46}   
\newcommand{\medianrefobs}{6500}             
\newcommand{\minrefobs}{4600}               

\newcommand{\pHickenS}{35}
\newcommand{\pHickenC}{4}

\newcommand{\minexp}{300}
\newcommand{\maxexp}{3600}

\newcommand{\medianexp}{1800}

\newcommand{\galacont}{40}
\newcommand{\galasl}{30}

\newcommand{\galbcont}{80}
\newcommand{\galbsl}{30}
\newcommand{\galbz}{0.132}

\newcommand{\slitid}{16287}
\newcommand{\slitspid}{1449}
\newcommand{\slitlcage}{2}

\newcommand{\contfilter}{$g$}
 
\begin{document}

   \title{NTT and NOT spectroscopy of SDSS-II supernovae}

   \author{L. \"{O}stman\inst{1,2,3}, J. Nordin\inst{1,3}, A. Goobar\inst{1,3}, R. Amanullah\inst{1,3}, M. Smith\inst{4,5}, J. Sollerman\inst{3,6,7}, V. Stanishev\inst{8}, M.~D. Stritzinger\inst{3,7,9}, 
B.~A. Bassett\inst{5,10,11}, T.~M. Davis\inst{7,12}, E. Edmondson\inst{4}, J. A. Frieman\inst{13,14}, P.~M. Garnavich\inst{15}, H. Lampeitl\inst{4}, G. Leloudas\inst{7}, J. Marriner\inst{13}, R.~C. Nichol\inst{4} , K. Romer\inst{16}, M. Sako\inst{17}, D.~P. Schneider\inst{18}, C. Zheng\inst{19}
}


\institute{
Department of Physics, Stockholm University, 106 91 Stockholm, Sweden.  
\and 
Institut de F\'{i}sica d'Altes Energies, 08193 Bellaterra, Barcelona, Spain. 
\and
Oskar Klein Centre for Cosmo Particle Physics, AlbaNova University Center, 106 91 Stockholm, Sweden.    
\and
Institute of Cosmology and Gravitation, Portsmouth PO13FX, United Kingdom. 
\and
Department of Mathematics and Applied Mathematics, University of Cape Town, South Africa. 
\and
Astronomy Department, Stockholm University, 106 91 Stockholm, Sweden. 
\and
Dark Cosmology Centre, Niels Bohr Institute, University of Copenhagen, 2100 Copenhagen \O, Denmark.  
\and
CENTRA - Centro Multidisciplinar de Astrof\'{i}sica, Instituto Superior T\'{e}cnico, 1049-001 Lisbon, Portugal. 
\and
Carnegie Institute for Science, Carnegie Observatories, Casilla 601, La Serena, Chile. 
\and
South African Astronomical Observatory, Cape Town, South Africa. 
\and
African Institute for Mathematical Sciences, Muizenberg, Cape Town, South Africa. 
\and
School of Mathematics and Physics, University of Queensland, QLD 4072, Australia 
\and
Center for Particle Astrophysics, Fermi National Accelerator Laboratory, Batavia, Illinois 60510, USA. 
\and 
Kavli Institute for Cosmological Physics, University of Chicago, Chicago, Illinois 60637, USA. 
\and
University of Notre Dame, Notre Dame, IN 46556-5670, USA. 
\and
Department of Physics and Astronomy, University of Sussex, UK.    
\and
Department of Physics and Astronomy, University of Pennsylvania, Philadelphia, PA 19104, USA. 
\and
Department of Astronomy and Astrophysics, Pennsylvania State University, University Park, PA 16802 USA. 
\and
Kavli Institute for Particle Astrophysics and Cosmology, Stanford University, Stanford, CA 94305-4060, USA. 
}

   \date{Received 7 September 2010; accepted 16 October 2010}

  \abstract
{The Sloan Digital Sky Survey II (SDSS-II) Supernova Survey, conducted
  between 2005 and 2007, was designed to detect a large number 
  of Type Ia supernovae around $z\sim0.2$, the redshift ``gap'' 
  between low-$z$ and high-$z$ supernova searches.
 The survey has provided multi-band ($ugriz$)
  photometric lightcurves for variable targets, and supernova candidates
  were scheduled for spectroscopic observations, primarily to 
  provide supernova classification and accurate redshifts. We present
  supernova spectra obtained in 2006
  and 2007 using the New Technology Telescope (NTT) and the Nordic
  Optical Telescope (NOT).}
%
{We provide an atlas of supernova spectra in the range $z=${\zminiacrop}--{\zmaxiacrop} 
that complements the well-sampled lightcurves from SDSS-II 
in the forthcoming three-year SDSS supernova cosmology analysis. The sample can, for example, be used for spectral studies of Type Ia supernovae, which are critical for understanding potential systematic effects when supernovae are used to determine cosmological distances.}
%
{The spectra were reduced in a uniform manner, and special care was
taken in estimating the uncertainties for the different processing steps. Host-galaxy light was 
subtracted when possible and the supernova type fitted using the SuperNova IDentification code (SNID). We also
present comparisons between spectral and photometric dating using SALT lightcurve fits to the photometry from SDSS-II, 
as well as the global distribution of our sample in terms of the lightcurve parameters: stretch and colour.}
%
{We report new spectroscopic data from {\nosneia} Type Ia supernovae, mainly between $-$9 and $+$15 days from lightcurve maximum, including a few cases of multi-epoch observations. This homogeneous, host-galaxy subtracted, Type Ia supernova spectroscopic sample is among the largest such {\dataset}s and unique in its redshift interval. 
The sample includes two potential SN 1991T-like supernovae (SN 2006on and SN 2007ni) and one potential SN 2002cx-like supernova (SN 2007ie).
In addition, the new compilation includes spectra from {\nosneII} confirmed Type~II and {\nosneIbc} Type~Ib/c supernovae.}
{}


   \authorrunning{\"{O}stman, Nordin et al.}

   \maketitle


\section{Introduction}
\label{sec:intro}

Type Ia supernovae (SNe~Ia) as distance indicators provided the first
direct evidence of the late-time acceleration of the Universe
\citep{1999ApJ...517..565P,1998AJ....116.1009R}. The observed faintness
of high-$z$ SNe~Ia suggests there is an energy
component with a negative pressure, which has been given the name \emph{dark energy}. The
origin and nature of the dark energy is still unknown, and precise
measurements of its equation of state
remain as one of the key goals for both cosmology and fundamental
physics. To achieve the necessary precision using SNe~Ia, large
statistical samples and excellent control of potential systematic
effects are imperative.

The Sloan Digital Sky Survey II (SDSS-II) Supernova Survey
\citep{2000AJ....120.1579Y,2008AJ....135..338F} operated as a three-year 
(3 months per campaign) program (2005-2007), aimed at detecting 
a significant number of intermediate-redshift SNe~Ia in a rolling survey. 
The survey provided multi-band ($ugriz$) photometric lightcurves for transient
targets. The SN candidates detected in the galaxy subtraction pipeline were
scheduled for spectroscopic observation, in order to provide 
spectral identification and an accurate 
redshift. SN~Ia candidates were given highest priority for spectroscopic follow-up, 
however other SN types were also observed.

The first-year photometry and spectroscopy have been presented in
\citet{2008AJ....136.2306H} and \citet{zheng08}, respectively.
First cosmological results, including a Hubble diagram
consisting of 103 SNe~Ia discovered during the first year of the survey can be found in
\citet{2009ApJS..185...32K} \citep[see also][]{2009ApJ...703.1374S,2010MNRAS.401.2331L}.

In this article we present optical spectroscopy obtained with the ESO
New Technology Telescope (NTT) and the Nordic Optical Telescope
(NOT). In total, {\nospectra} spectra of SDSS-II SN candidates were obtained during 2006 and 2007. 
The {\dataset} contains {\nosniaspectra} confirmed SN~Ia spectra 
from {\nosneia} objects in the redshift range {\zminiacrop} $<$ $z$ $<$ {\zmaxiacrop}. The redshift range complements previous and current SN~Ia surveys and the spectra are, in general, of high signal-to-noise ratio
(S/N).

The NTT and the NOT spectra constitute a natural subset of the
complete SDSS sample since many of the observers were active at both
telescopes guaranteeing a consistent observing strategy as well as
similar choices of targets. Furthermore, all data were processed using the same pipeline, in a coherent manner including a detailed error analysis.
The reduction procedure is described in this paper. The {\dataset} is presented as a documented 
library available for future studies of SNe~Ia.

When our SN~Ia spectra are combined with low-redshift
spectroscopic samples, they cover a wide interval in cosmic time.
They can then be used to search for signs of evolution of luminosity with distance, a potentially
large systematic uncertainty in the use of SNe Ia as distance indicators
\citep[see e.g.,][]{2008JCAP...02..008N}.
A statistical study of spectral features of SNe~Ia could provide
evidence for or against such evolution. In order to make such a study possible, correct estimates of 
the uncertainties are essential, especially if comparisons are to be made to
a high-S/N local sample. %
In a companion paper, \citet{nordin10} present initial results of such a quantitative study.
Similar studies, using {\dataset}s in other redshift ranges have previously been made \citep{2005AJ....130.2788H,2006AJ....131.1648B,2007A&A...470..411G,2008ApJ...684...68F,2008A&A...477..717B}.
Furthermore, spectroscopic features can potentially be used as brightness indicators \citep{2008A&A...477..717B}.
Spectroscopic studies of SNe also play an important role to constrain 
the progenitor systems, and the explosion physics.

Other large spectroscopic samples which have been made accessible to the community include the low-redshift sample of \citet{2008AJ....135.1598M} (432 spectra of 32 SNe Ia) and the high-redshift sample of \citet{2009A&A...507...85B} (139 ESO/VLT spectra of 124 SNe Ia from SNLS with an average redshift of 0.63). The NTT/NOT {\dataset} complements these two well, having a mean redshift of {\zmean} and a similar number of spectra as the SNLS spectra. The S/N of our {\dataset} is lower than the low-redshift sample and higher than the SNLS sample.
Descriptions of other medium- and high-redshift samples have been presented in e.g. \citet{2005ApJ...634.1190H,2009AJ....137.3731F,zheng08}.
 
The NTT/NOT sample also contains spectra of core-collapse SNe. Some of the SNe IIP presented in this paper were used by \citet{2010ApJ...708..661D} to investigate the method where luminosity is standardised using the ejecta velocity during plateau phase.

The paper is organised in the following manner. Section~\ref{sec:obs} describes the spectroscopic observations performed with the NTT and the NOT. In Section~\ref{sec:data} the
{\dataset} is presented and comparisons with other {\dataset}s are made.
Section~\ref{sec:reductions} describes the reduction method. The problems with differential atmospheric refraction and slit losses are addressed in Section~\ref{sec:dar}.
Section~\ref{sec:host} deals with the host-galaxy subtraction and Section~\ref{sec:typing} with the typing. A couple of specific objects are briefly discussed in Section~\ref{sec:special}.
This is followed by a summary in Section~\ref{sec:results}. The appendix contains three
tables describing the spectra: the observations (\ref{tab:obs}), the
typing and redshifts (\ref{tab:type}) and the quality of the data
(\ref{tab:qual}).

%

\section{Observations}
\label{sec:obs}

The NTT, located at the La Silla Observatory in Chile, was used for spectroscopic
observations of SDSS-II SN candidates from September to December in
2006 and 2007.\footnote{The observations were acquired in the ESO programmes
077.A-0437, 078.A-0325, 079.A-0715 and 080.A-0024 under PI Robert
Nichol.}
Thirty-four nights were awarded for the project, out of which
{\nttnights} had sufficiently good conditions to obtain SN spectra. 
Through the course of these nights, {\noobsntt} spectra of SN candidates were obtained. 
The primary mirror of the NTT has a diameter of 3.58-m. 
The observations were performed using the ESO Multi-Mode Instrument
(EMMI; \citet{1986SPIE..627..339D}) in the Red Imaging and
Low-Dispersion spectroscopy (RILD) mode using grism 2. 
Grism 2 provides a wavelength coverage from 3800 to 9200 {\AA}. It has 300 grooves per mm, a wavelength dispersion of 1.74 {\AA} per pixel, and a spatial resolution of
0$\farcs$166 per pixel, without CCD binning. During the observations a binning of 2$\times$2 was used, resulting in 
a resolving power $R\simeq$570 at 6000 {\AA} for a 1$\farcs$0 slit.
Slit widths of 1$\farcs$0 or 1$\farcs$5 were used depending on the seeing conditions.

The NOT is located at the Observatorio del Roque de los Muchachos on La Palma, Spain. 
Spectroscopic observations were conducted with this telescope during November 2006 and September and
November 2007.\footnote{The observations were acquired in the programmes with proposal numbers 34-004,
35-023 and 36-010 under PI Maximilian Stritzinger.}
Eleven nights were awarded for the project, out of which {\notnights} nights had good enough observing conditions to obtain SN spectroscopy. During these nights {\noobsnot} spectra of target SNe were obtained. 
The primary mirror of the NOT has a diameter of 2.56-m. 
Spectra were obtained with the Andalucia Faint Object Spectrograph and Camera
(ALFOSC) using grism 4, which has 300 grooves per mm. This set-up provides 
a wavelength range from
3200 to 9100 {\AA} with a wavelength dispersion of
$\sim$3.0 {\AA} per pixel and a spatial resolution of 0$\farcs$19 
per pixel. The resolving power is 710 for a 0$\farcs$5 slit.
Depending on the seeing, slit widths of 1$\farcs$0 or 1$\farcs$3 were used.

In general, to avoid observing erroneous objects such as asteroids and 
AGNs, each candidate had at least two epochs of photometry before they
were placed in our spectroscopic queue.
High priority was given to probable SNe~Ia which
appeared to have been discovered while still being on the rise.
Furthermore, since several telescopes were taking spectra on the same
nights, a division was made where faint targets were typically allocated 
to telescopes with a larger aperture. More details about the procedure
for selecting the objects for spectroscopic observation can be found in
\citet{2008AJ....135..348S}.

A typical exposure time for the spectroscopic observations was {\medianexp} seconds, but depending on the magnitude of the SN and the observing conditions, exposure times were varied between {\minexp} and {\maxexp} seconds. In Table~\ref{tab:obs} the exposure time for each individual spectrum is given.

As part of the observing programme, host-galaxy spectra of previously
confirmed SNe~Ia were obtained when 
SN candidates were lacking or the observational conditions were poor. These host galaxy spectra 
will not be presented in this paper.


\section{{\Dataset}}
\label{sec:data}

\begin{figure}
	\centering
	\includegraphics[width=8cm]{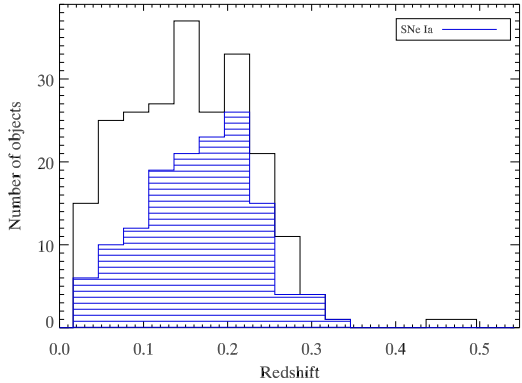}
	\caption{Redshift distribution of the {\nozhist} observed
	objects at the NTT and the NOT with a known redshift. The subset of
	objects classified as SNe~Ia is shown in the striped
	histogram. The bin size in the plot is {\zBins}.}
	\label{fig:zdistr}
\end{figure}

The {\dataset} contains {\nospectra} spectra of {\noobj} individual objects.
Information about the spectra are given in the tables in the appendix. Each object has been given a unique SDSS ID and the spectra have been given unique spectral numbers, SPIDs.
Redshifts were obtained from Zheng et al (in preparation). Out of the {\noobj} targets, {\nozsdss} had prior galaxy redshifts from SDSS DR7, {\nozgal} objects had the redshift determined
from galaxy lines and {\nozsn} from SN features. Some of these redshifts were determined from NTT/NOT spectra, but also observations from other telescopes were used. Furthermore, there were {\nozsurveytext} objects ({\snid{{\surveyz}}}) for which no reliable redshifts could be obtained from either the host galaxy or SN features, but where a redshift could be determined through a rough template fitting. 
For {\noobjnoz} objects, no redshift could be determined at all. Most of these objects are seemingly ``hostless'' events, several exhibiting lightcurves which differ significantly
from that of SNe.
The redshifts and the origin of the redshifts are listed in Table~\ref{tab:type}.
The redshift distribution of the objects is shown in
Figure~\ref{fig:zdistr}, where the subset of objects which were
classified as SNe~Ia is indicated.
The redshift range for all objects observed was
$z=$ \ {\zminall}$-${\zmaxall}, and for the subset of SNe~Ia,
$z_{SNIa}=$ \ {\zminia}$-${\zmaxia}.

Our SN~Ia sample contains {\nosniaspectra} spectra 
of {\nosneia} individual objects, plus {\nosneiaq} likely SNe Ia.
In addition, we also obtained {\nosnIIspectra} spectra of {\nosneII} SNe~II 
and {\nosnIbcspectra} spectra of {\nosneIbc} SNe~Ib/c.
{\nogalspectext} spectra were identified as galaxy
spectra, which presumably were the result of poor seeing conditions 
or that the spectra were taken when the SN
had faded significantly, becoming much fainter than the host galaxy.
These spectra are useful as they provide spectroscopic redshift determination for the 
photometric sample of SNe. 
In Section~\ref{sec:typing} we describe how the spectroscopic typing was performed.
The number of spectra and the number of unique objects are summarised in Table~\ref{tab:numbers}.
\begin{table}
\begin{center}
\caption{\label{tab:numbers} Number of spectra and unique objects.}
\begin{tabular}{lll}
\hline \hline
& Spectra & Objects \\
\hline
Total & {\nospectra} & {\noobj} \\
~ with redshift & ~ {\nospecwz} & ~ {\noobjwz} \\
\hline
SN~Ia & {\nosniaspectra} & {\nosneia} \\
~ with good LC & ~ {\nosniaspectralc} & ~ {\nosneialc}\\
SN~Ia? & {\nosniaspectraq} & {\nosneiaq} \\
SN~II & {\nosnIIspectra} & {\nosneII} \\
SN Ib/c & {\nosnIbcspectra} & {\nosneIbc} \\
Not SN & {\nonotsnspec} & {\nonotsn} \\
~ Galaxy & ~ {\nogalspec} & ~  {\nogal} \\
Unclassified & {\nounknownspec} & {\nounknown} \\
\hline
\end{tabular}
\end{center}
\end{table}
For the first four years of ESSENCE, 55\% of the spectroscopically observed objects were identified as SNe Ia, probable SNe Ia or core collapse supernovae \citep{2009AJ....137.3731F}.
The corresponding number for the NTT/NOT {\dataset} is {\snproc}\%. The SN fraction obtained depends on the design of the search, the redshift interval, the size of the telescope and the time spent on each supernova target. As two of several telescopes (with different sizes) involved in the follow-up of SDSS SNe, where different targets were assigned to different telescopes, the evaluation of the efficiency is further complicated.

\begin{figure}
	\centering
	\includegraphics[width=8cm]{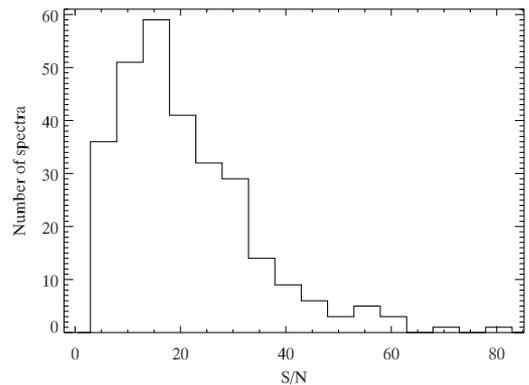}
	\caption{Distribution of the S/N for the spectra
	taken at the NTT and the NOT. The S/N is calculated in 10 {\AA} bins and averaged over the interval between 4000 and 6500 {\AA} (in observed frame).}
	\label{fig:ston}
\end{figure}

The spectroscopic S/N distribution for the {\nospectra}
NTT and NOT spectra is shown in Figure~\ref{fig:ston}. Values are computed for 10 {\AA} bins and then averaged over the interval between 4000 and 6500 {\AA} (in observed frame), which is the interval of most interest for SN features.

%

\subsection{Lightcurve properties}
\label{sec:phot}

SDSS multi-band photometry of the objects in our sample and their host
galaxies were obtained from $ugriz$ \citep{1996AJ....111.1748F} observations using the method presented in \citet{2008AJ....136.2306H}.
Lightcurves of all confirmed SNe~Ia were fitted using SALT
\citep{2005A&A...443..781G}. 
The same authors have since published a new fitter, SALT2 \citep{2007A&A...466...11G}. Even though SALT2 in many respects is an improvement, we use the older code since it provides smooth lightcurves which are easy to use for interpolation, e.g. for calculating the photometry of the spectral epoch, to estimate the host galaxy contamination.
The fitted parameters are: lightcurve
width $s$ (``stretch''), SALT colour $c$, rest-frame peak
$B$-band magnitude, and time of $B$-band maximum light. 
However, {\nosneiabadlc} SN~Ia lightcurves lack
either pre- or post-maximum photometry. 
This leads to bigger uncertainties for the derived
lightcurve parameters. In the remainder of
this paper, we will refer to these as \emph{poor} lightcurves. Removing all SNe~Ia with poor lightcurves leaves
{\nosniaspectralc} spectra of {\nosneialc} different SNe~Ia.

\begin{figure}
	\centering
		\includegraphics[width=8cm]{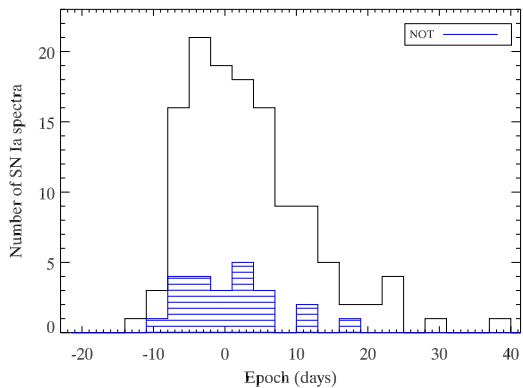}
	\caption{Epoch distribution of the SN~Ia spectra observed at
	the NTT and the NOT, excluding the SNe with poor lightcurves. The
	un-filled histogram shows the full sample of SN Ia spectra, while the
	striped histogram shows the subset observed at the NOT. The epoch is
	defined as the number of days in rest frame since $B$-band
	maximum brightness, as obtained from the
	lightcurve. The {\noephist} spectra are divided into epoch
	bins with a width of {\epochBins} days.}
	\label{fig:hist_epoch}
\end{figure}

The epochs of the SN~Ia spectra, calculated as the number of days in rest
frame from $B$-band maximum light, are listed in Table~\ref{tab:type}, and
the sample distribution is shown in Figure~\ref{fig:hist_epoch}. 
In the table all epochs calculated from poor lightcurves are marked with a superscript
'$p$'. The uncertainty in the epoch is obtained as either the uncertainty in peak date obtained from SALT or through error simulations, whichever is bigger. The error simulations were performed by randomising the photometry within the error bars and making new SALT fits, after which the spread in the peak date was studied. The epoch errors include the uncertainty in the photometry, but they do not include any potential systematic uncertainties in the SALT fitting procedure.

\begin{figure}
	\centering \includegraphics[width=8cm]{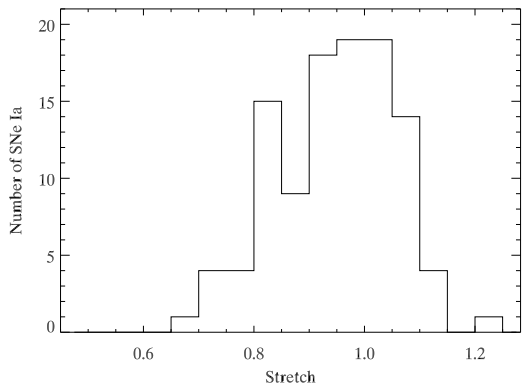}
	\caption{Distribution of stretch (SALT $s$ parameter) of the SNe~Ia observed at the NTT
	and the NOT, excluding the SNe with poor lightcurves. The mean value is {\sMean} and
	the standard deviation {\sStd}. The {\noshist} objects are
	divided into bins of size {\sBins}. Converting the stretch to
	$\Delta m_{15}(B)$ following \citet{2006AJ....131..527J}, a
	mean value of {\deltamMean} and a standard deviation of
	{\deltamStd} is obtained.}
	\label{fig:hist_stretch}
\end{figure}
The stretch distribution, shown in Figure~\ref{fig:hist_stretch}, is similar to the distribution of the Constitution set \citep{2009ApJ...700..331H}, both regarding the mean value and the shape. The Kolmogorov-Smirnov (K-S)
test gives a probability of {\pHickenS}\%, which is strong evidence against rejecting the assumption that the two samples of stretch belong to the same distribution.
\begin{figure}
	\centering \includegraphics[width=8cm]{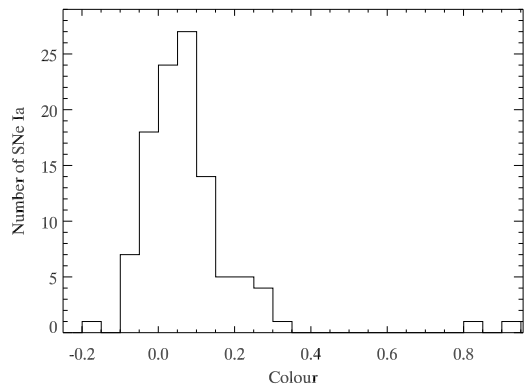}
	\caption{Distribution of colour (SALT $c$ parameter) for the
	SNe~Ia observed at the NTT and the NOT, excluding the SNe with poor lightcurves. The mean value is {\cMean} and the median
	{\cMedian}. The standard deviation is {\cStd}. The {\nochist}
	objects are divided into bins of size {\cBins}.}
	\label{fig:hist_colour}
\end{figure}
The distribution of colour (SALT $c$ parameter), 
presented in Figure~\ref{fig:hist_colour},
has a K-S probability of {\pHickenC}\% when compared to the distribution of the Constitution sample of
\citet{2009ApJ...700..331H}.
We conclude that the lightcurve properties of the SNe~Ia in our sample
are statistically compatible with the SN lightcurves previously used in 
cosmological analyses.

%

\subsection{Data archive}
\label{sec:archieve}

The spectra are publicly available in electronic format.\footnote{\texttt{http://www.physto.se/}$\sim$\texttt{linda/spectra/nttnot.html}}
In addition to the calibrated spectra, a corresponding error spectrum is also provided.
Versions of the spectra are available both with and without host-galaxy subtraction.
Additional information about the spectra on the website; observing conditions, object type, days since maximum brightness, redshift, etc are available in Appendix~\ref{sec:tables} of this paper.

%

\section{Data reduction}
\label{sec:reductions}

The spectroscopic data were reduced using the Image Reduction and Analysis Facility
({\iraf})\footnote{{\iraf} is distributed by the National Optical
Astronomy Observatories, which are operated by the Association of
Universities for Research in Astronomy, Inc., under cooperative
agreement with the National Science Foundation.} and our own IDL
routines. Below follows a step-by-step description of the reduction
procedure.


\paragraph{Bias subtraction}

Bias frames were taken on every night. No spatial variations were
detected so we used the CCD overscan region to subtract the
bias.


\paragraph{Flat fielding}

\begin{figure}
	\centering
		\includegraphics[width=4cm]{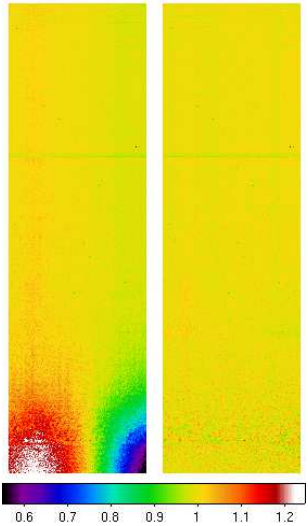}
	\caption{A normalised flat field from the NTT before
	({\em left}) and after ({\em right}) the correction for the zeroth order
	light was applied. The figure shows only one
	of the two chips of the CCD. The vertical direction is the dispersion axis and the horizontal direction
	is the spatial axis. The scales are the same
	for the two images.}
	\label{fig:flat}
\end{figure}

The flat fields from the NTT observations had a peculiar feature at
the blue end of the spectrum corresponding to wavelengths 
shorter than roughly 5200 {\AA} (see the left panel of
Figure~\ref{fig:flat}). This is an effect from the zeroth order image of the
grism which, while being outside the CCD area, still produced a visible glow on the detector.
The behaviour was not present in any of the other frames and was 
removed from the flat fields by fitting a surface over the
region.
In the right panel of Figure~\ref{fig:flat}, the corrected and
normalised flat is shown. Some horizontal lines can be seen in the flat
field image around 7600 {\AA}. These are absorption lines due to water
vapour and molecular oxygen, and are caused by the long light path
inside the instrument. This leads to a variation of the order of 3.5\% in the normalised
flat.\footnote{See the documentation about EMMI \texttt{www.eso.org/sci/facilities/lasilla/instruments/emmi}.}
 
For a given night of observation, one flat field was constructed for each grism/slit-width combination, which 
was then used to flat field the 2-D spectral images of the standard stars and the SN candidates.


\paragraph{Spectral extraction}

The SN spectra were extracted using the optimal extraction
algorithm of \citet{1986PASP...98..609H}.
An extraction window was chosen in the spatial direction for a narrow
wavelength range and was then traced for all wavelengths.
In the cases when the SN was not well separated from the host
galaxy, a small spatial extraction window was used to minimise 
contamination from host-galaxy light.
For the sky background subtraction, two small spatial ranges were
defined on each side of the SN, and a fitted linear function was
subtracted. For extended host galaxies,
when the SN light was separated from the core of the host galaxy, this
background fit also included host-galaxy light. Any residual host-galaxy
light present in the spectra was removed at a later stage in the
reduction process (see Section~\ref{sec:hostsub}).
When the combination of SN and host galaxy was complex, we
experimented with different sizes of extraction apertures and higher
orders of background fits, but without any significant improvement in
the final result.


\paragraph{Wavelength calibration}

For wavelength calibration, spectra of a helium-argon lamp were
taken at the NTT and of a helium lamp at the NOT.
For each object spectrum, an arc spectrum was extracted with the same
centring and trace as the object spectrum. 
A Chebyshev polynomial of fifth or sixth order was fitted to the
identified wavelengths for the arc lamp emission lines. The
solution was checked against locations of sky lines.
The wavelength solution was then applied to the object
spectrum.


\paragraph{Correction for second order contamination}

When obtaining spectra over a wide wavelength range in one single
exposure, contamination from second order diffraction could 
lead to an erroneous flux where there is an overlap of the orders. 
The second order contamination can be circumvented by using a blocking
filter for blue light or by ignoring the red end of the
spectrum. However, this reduces the wavelength range of the
spectrum. Another method is to double the exposure time and divide the
observations into a blue part and a red part. However, for faint
objects which require long exposure times, this is not desirable.
A method has been developed by \citet{stanishev07} to correct spectra,
obtained at the NOT using grism 4, for second order contamination during
the reduction phase. With this method the full spectrum can be used.
Observations were made at the NTT using grism 2 to derive the necessary information to correct
spectra from that telescope too.
To be able to correct for the second order contamination we need to
(1) find out the wavelength relation between the first-order spectrum
and the second-order spectrum
and (2) find the ratio of the efficiencies of the grism for the two orders.
The wavelength overlap of the two orders were determined using observations of arc lamps, together with blocking filters.
To determine the flux relation between the two orders, bright blue
stars were observed with and without order-blocking filters.
\citet{szokoly04} have developed a similar method.

\begin{figure}
	\centering
		\includegraphics[width=8cm]{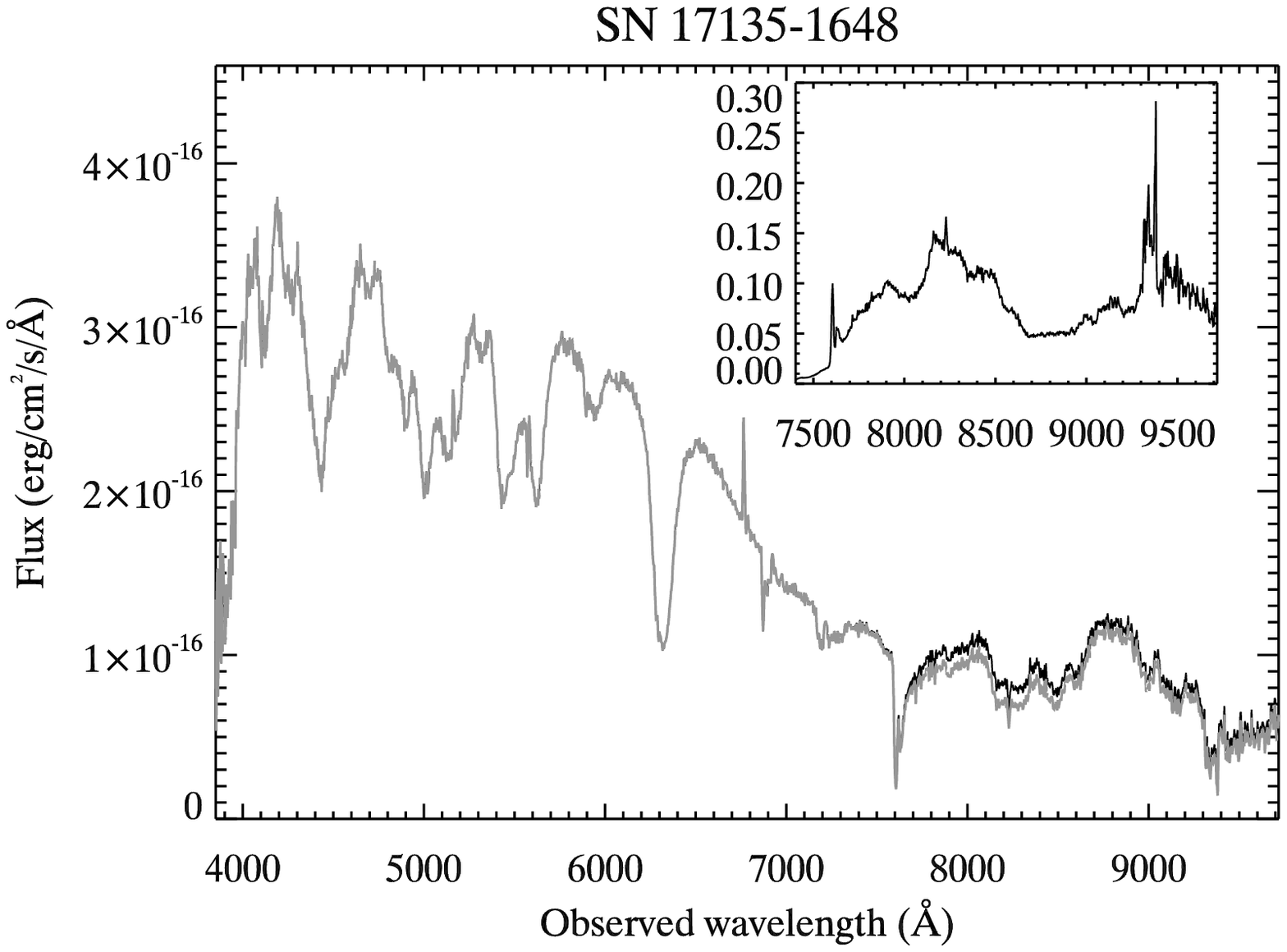}
		\includegraphics[width=8cm]{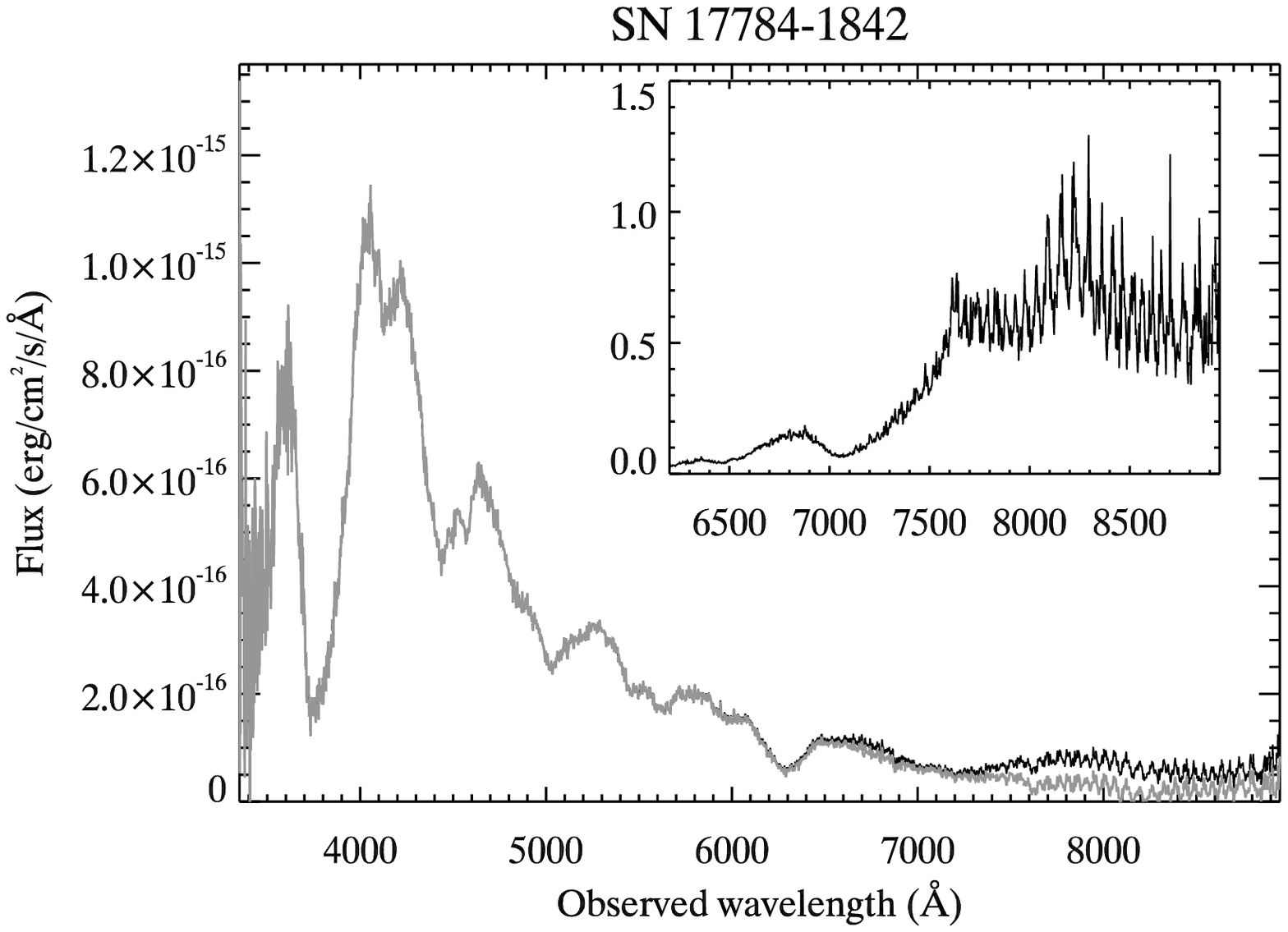}
	\caption{Second order corrections of two SNe, 
{\snid{17135}} (SN 2006qm) observed at the NTT (top panel) and {\snid{17784}} (SN 2007jg)
	observed at the NOT (bottom panel). The observed flux is displayed in black and the
	corrected flux in light grey. The spectra have been flux
	calibrated. No corrections due to telluric
	absorption, Milky Way extinction or effects from host-galaxy
	light have been applied to the spectra shown here. The inset in each panel shows the relative correction with respect to the observed flux.}
	\label{fig:second}
\end{figure}

In extreme cases the second order contamination can be as strong as the flux at long wavelengths. The size and shape of absorption lines at these wavelengths can also be
affected. This is illustrated in Figure~\ref{fig:second}, where two examples of second order corrected spectra are shown, one for each 
telescope.
If the second order contamination is not removed, systematic
effects could be introduced since blue SNe are more affected by
the contamination than red SNe.

%

\paragraph{Flux calibration}

To correct the object spectra for the wavelength dependent sensitivity
of the detector, the spectra were flux calibrated using
spectrophotometric standard-stars.
For the NTT, the standard stars were observed with a slit width of
 5$\farcs$0 and for the NOT with a slit of width 1$\farcs$3. A wider slit
 width is used to make sure that the seeing does not affect the flux
 through the slit. All standard stars were
 observed at parallactic angle, to avoid effects from differential
 atmospheric diffraction.
%
%
The standard-star spectra were processed in the same way as the object
spectra.

Using {\iraf}, a sensitivity function was created for each instrument,
slit width and observing session (of up to five nights). For the observations at the NOT during September and
November 2007, a common sensitivity function was created since few
standard star observations could be used. No significant variations with time were seen.

Corrections due to atmospheric extinction were implemented for La Silla and 
La Palma.

For a significant fraction of the SN observations, the parallactic angle was not used, causing differential slit losses due to atmospheric refraction. This problem will be discussed in detail in Section~\ref{sec:dar}.

%

\paragraph{Telluric removal}

The spectra have been corrected for telluric absorption. Using
standard-star observations the telluric absorptions were isolated and
then combined into one telluric spectrum for each slit width and night
of observations. The SN spectra were then corrected through
division of a scaled telluric spectrum. The scale factor was
determined as the optimal weight for removing any telluric absorption
near 7600 {\AA}. This is the dominant telluric region, while not
containing any interesting SN features for objects in our redshift range.
In some cases no weight could be found such that the telluric spectrum
obtained during the same night could remove the telluric feature. If
another telluric spectrum obtained in close connection to the
SN spectrum could do the matching, this was used instead. Some spectra could not be corrected perfectly using any telluric spectrum.
The increased uncertainty in the spectra due to possible un-corrected telluric absorption was estimated by calculating the deviation from a spline fit
through the 7600 {\AA} region, after correction. This error was then added to the SN error-spectrum.

%

\paragraph{Correction for dust extinction in the Milky Way}

The spectra were corrected for dust extinction in the Milky Way using
$R_V=3.1$ and the colour excess $E(B-V)$ from
\citet{1998ApJ...500..525S}. The extinction was assumed to follow the
\citet{cardelli89} extinction law modified according to \citet{odonnell94}.
According to \citet{1998ApJ...500..525S}, the uncertainty on $E(B-V)$
is of the order of 16\%. The errors due to the uncertainty in $R_V$
and the uncertainty in the extinction law was found to be of less
importance. The wavelength-dependent errors from the Milky Way
correction were added in quadrature to the existing error
spectrum.

%

\section{Differential atmospheric refraction and slit losses}
\label{sec:dar}

\emph{Differential atmospheric refraction} (DAR) occurs when an observed image is spread out with wavelength along the direction toward the horizon due to the wavelength dependence of the refraction index of air \citep[see e.g.][]{filippenko82,1998SPIE.3355...36C}.
As a result, objects centred on the slit for one wavelength, may be
partially or even fully outside the slit for shorter and longer
wavelengths. 
The effect can be minimised by (1) observing at a low airmass and (2) rotating
the slit to the parallactic angle, where the slit is parallel to the
direction of the atmospheric refraction, i.e. normal to the horizon.
Other factors affecting the differential slit loss are:
slit width, seeing, morphology of the observed object, spectral
coverage and the wavelength(s) for which the object is centred.

\begin{figure}
	\centering
	\includegraphics[width=8cm]{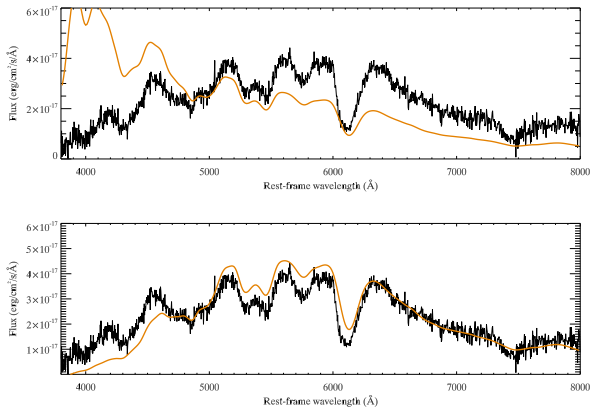}
	\caption{The observed spectrum of {\snid{{\slitid}}} (SPID {\slitspid}) {\slitlcage} days past maximum brightness is shown in black together with the Hsiao template at the same epoch in orange (upper panel). In the lower panel the Hsiao template has been multiplied with the theoretically calculated differential loss due to the atmospheric refraction.}
	\label{fig:sl}
\end{figure}

The spectroscopic observations of SNe presented in this paper were typically
{\em not} made at parallactic angle.\footnote{The observations of the standard stars were always made at parallactic angle.} Instead, the slit orientation was
most often chosen to simultaneously give a spectrum of the host galaxy.
The differential atmospheric refraction (in arcseconds) for a certain
wavelength with respect to a reference wavelength was estimated using
the formulae by \citet{1967ApOpt...6...51O}. The reference wavelength
is the wavelength with which the SN had been centred on the
slit.
To calculate the fraction of the flux within the slit for each wavelength we
modelled the SN flux with a Gaussian distribution using the
seeing at the time of observations.
In Figure~\ref{fig:sl}, we show one of the observed spectra with an insignificant host-galaxy contamination and a severe differential slit loss. The spectrum is shown together with a SN Ia template from \citet{2007ApJ...663.1187H} (which we from now on will refer to as Hsiao templates) in its original form as well as affected by the estimated differential slit loss. The slit loss affected template is a \emph{much} better fit to the observed spectrum. The estimated slit loss at very short wavelengths is exaggerated, which seems to be true for a large fraction of the theoretically calculated slit loss functions. In most cases such a direct comparison cannot be made due to the complex interplay of host-galaxy contamination and slit loss effects.
In Table~\ref{tab:qual} we indicate the estimated effect from the
differential atmospheric refraction on the spectra at an observed
wavelength of 4000 {\AA}.
It should be noted that this is only a rough estimate of the
differential slit loss since many simplifications are used in the
calculations and the centring over the slit was not
perfect. 
For a non-negligible fraction of the NTT/NOT spectra, the differential slit loss is severe at short wavelengths. This is due to large airmasses and large differences between the observing angle and the parallactic angle. The median value of the airmass of the
observations is {\medianairmassobs}, with {\perchighairmassobs}\% of
the observations having an airmass larger than 1.5. Large airmasses were unavoidable because of the position of the
SDSS SN fields relative to the NTT and the NOT sites, during the observing season.
Deviations from observations at the parallactic angle were also fairly large: nearly half of the spectra ({\perchighangleobs}\%) were observed with 45$^{\circ}$ or more from the parallactic angle. The centring of the objects on the slit was done at wavelengths above {\minrefobs} {\AA}, typically at {\medianrefobs}
{\AA}.

A polynomial multiplier is included in the host-galaxy subtraction to account for the differential slit loss effects. However, these empirically determined corrections differ from the theoretically estimated functions described above since the polynomial also accounts for reddening.

%

\section{Host-galaxy contamination}
\label{sec:host}

A large fraction of the obtained SN spectra are contaminated by galaxy light. We present below an estimation of the galaxy contamination in the spectra from photometry and a description of the host-galaxy subtraction method applied to the data.

\subsection{Estimating the degree of host-galaxy contamination}

The host-galaxy contamination in the observed spectra was estimated using
multi-band SDSS photometry. 
To obtain an estimate of the galaxy light in the slit, the measured
photometric surface brightness of the galaxy at the position of the
SN was scaled to the area defined by the slit width times the
width of the spectral extraction window.
The expected SN flux in the observed spectra was calculated using the
SDSS lightcurve, interpolated to the night the spectrum was taken. For
objects where good lightcurve fits were available, these were used for
the interpolation. 
The amount of SN flux through the slit was estimated by modelling the SN flux as a Gaussian distribution with the width determined from the seeing.
The seeing variations between the different filters were estimated following \citet{1987sdap.book.....S} ($\propto \lambda^{-0.2}$).
In Table~\ref{tab:qual}, the estimated contamination in the {\contfilter}-band (observed frame)
for each observed spectrum is listed.
The contamination values are rough estimates and they are more likely over estimated since we assume that all galaxies are extended over the whole extraction window.
Out of the spectra, {\perlowcont}\% have an estimated contamination less than {\prcontlim}\%. The cases where extrapolation of the lightcurve was necessary are marked with a superscript of '$e$' in the table. These values should be considered less
reliable.

\subsection{Host-galaxy subtraction}
\label{sec:hostsub}

In the past, several different techniques have been used to subtract the host-galaxy contribution from observed spectra, such as $\chi^2$ based template fitting using the spectrum of the actual host galaxy or templates, either varying the contamination level or estimating it from photometry \citep[e.g.][]{2005ApJ...634.1190H,2008ApJ...674...51E}. For the vast majority of the NTT/NOT spectra, no good quality spectra of the host galaxy at the location of the SN were accessible at the time of the work and thus a $\chi^2$ based fitting using a real galaxy spectrum was rejected. To model the galaxy, principal component analysis (PCA) was chosen over template spectra. A similar method has also been used by \citet{zheng08}. There are also host-galaxy subtraction methods which separate the two components, SN och galaxy, in the two dimensional spectrogram \citep{2008A&A...491..567B}. However, this requires higher resolution of the data than that of the NTT/NOT {\dataset}.

For the NTT/NOT {\dataset} a number of different methods to subtract host-galaxy light from spectra were explored. We found the most stable method to be a principal component analysis based subtraction of galaxy eigencomponent spectra. For a large fraction of the moderately contaminated spectra (10-70\% contamination in the $g$-band) we found this method to yield good results.

It should be noted that some part of the host-galaxy light, at least for host galaxies with larger spatial extent on the CCD, were removed during a linear background fit in the reduction of the spectra. 
For these spectra the host-galaxy contamination that was fitted thus were the \emph{residuals} from the reduction.

An additional difficulty in the host-galaxy subtraction was that a large fraction of the NTT/NOT 
spectra were not observed at parallactic angle (see
Section~\ref{sec:dar}), and were thus affected by a wavelength dependent flux loss.

In the PCA-based subtraction method described here, SN templates were used. A potential worry is that the usage of templates constructed from normal low-redshift SNe might affect the outcome of the subtraction. This concern and other concerns are addressed in Section~\ref{sec:dischostgal}.

\paragraph{PCA-based galaxy subtraction}

The spectral energy distribution (SED) of the host galaxy was estimated through minimising the
difference between the observed spectrum and a combination of a SN
template and a set of galaxy eigencomponent spectra. The minimisation can be described with the formula
\begin{equation}
f_{\rm{fit}}(\lambda) = a_0 s(\lambda) \cdot f_{\rm{SN}}(\lambda) + \sum_{i=1}^n a_i g_{i}(\lambda),
\end{equation} 
where $f_{\rm{SN}}$ is the SN template, $g_i$ the galaxy eigencomponent spectra, $s$ is a second degree polynomial and, $a_i$ weights which are fitted in the subtraction. 

The galaxy eigencomponent spectra were created in a PCA analysis of 170\,000 SDSS galaxy spectra \citep{2004AJ....128..585Y}. The three most dominant eigenspectra have been shown to describe 99\% of all galaxies.\footnote{Galaxy emission lines can have errors of up to 10\% when using only three eigencomponent spectra \citep{2004AJ....128..585Y}. However, for our purposes it is enough with a well described continuum.} The fit can be made arbitrarily complex through the inclusion of more
galaxy eigencomponent spectra (to the existing three), but this extension was not used in the final studies since the fits
did not improve significantly while the computational demands increase
drastically.
In the fit we included the constraint that the total galaxy flux must be positive.

For the SN template in the fit we tried all Hsiao templates with epochs
$\pm$5 days from the spectral epoch as obtained from the
photometric lightcurve. We also included templates of more peculiar SNe, SN~1991bg and SN~1991T \citep{2002PASP..114..803N}, to study how well the Hsiao templates worked and to, possibly, find peculiar SNe in our sample. None of our spectra were well fitted with the SN~1991bg template (there will be a bias against such objects due to their faintness). Some spectra had better fits with a SN~1991T template.
The difference in the features sizes between the subtraction with the Hsiao template and the SN~1991T template was small in these cases.

To the SN template a second degree polynomial was multiplied which was introduced to account for reddening (e.g. due to host galaxy dust extinction) and differential slit loss effects. During observations, the object was centred on the slit in the red part of the spectrum, and thus differential slit losses due to atmospheric refraction (see Section~\ref{sec:dar}), when present, predominantly affects the blue end (as would extinction).
The second degree polynomial was fixed to have $s\equiv1$ at the wavelength $\lambda_1=6600$
{\AA} and $s<1$ for all other wavelengths. The value of $\lambda_1$ was chosen to correspond to the wavelength where most
spectra were centred on the slit. The function describing the differential slit loss
is asymmetric around this wavelength, but since the fit during
host-galaxy subtraction was only made between 4000 {\AA} and 6000 {\AA}, 
the behaviour of the slit loss function at longer wavelengths did not affect the fit.
The polynomial was only multiplied with the SN SED, and not with that of the galaxy. The reason for this was that since galaxies are not point sources, they are less affected by differential slit losses. 
A separate polynomial could be modelled for the fitted galaxy, but this would be both hard to evaluate and too computationally demanding, considering the small effect.
A test was performed where the same polynomial was added both to the galaxy and the SN, but this did not improve the quality of the fits.

\begin{figure*}
	\centering
        \includegraphics[width=8cm]{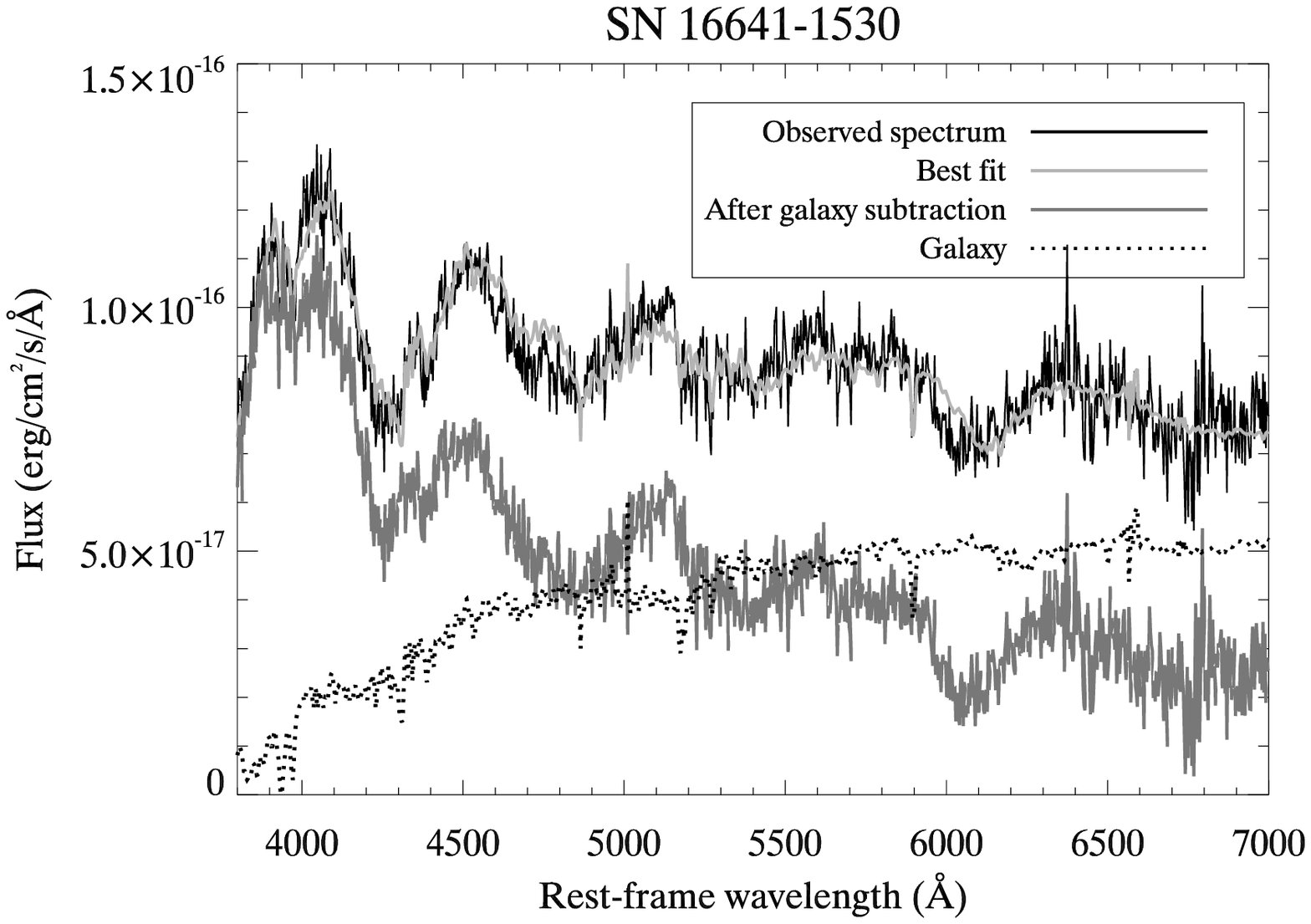}
        \includegraphics[width=8cm]{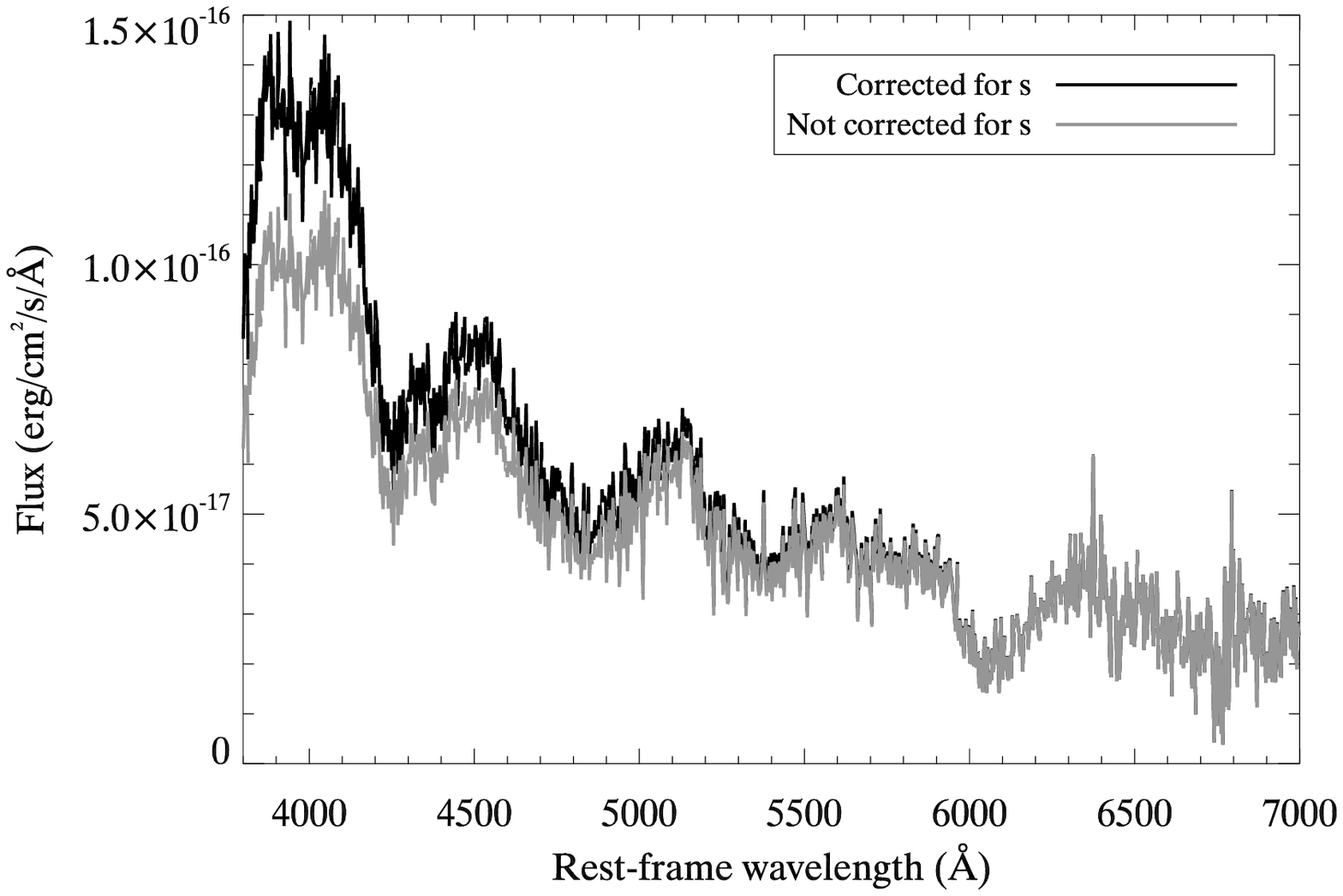}
       \caption{The PCA-based host-galaxy subtraction for {\snid{16641}} (SN 2006pr) at $z=$ {\galbz}. Left panel: The observed spectrum (black) is plotted together with the best fit (light grey). The galaxy template of the best fit is shown as a dotted line. The grey line shows the SN spectrum after subtraction of the estimated galaxy light. Right panel: The (host subtracted) SN spectrum is shown with and without correction for the wavelength dependent flux loss described by $s$ (black/grey). The host galaxy contamination estimated from photometry ({\galacont}\%) agrees well with the fraction of host galaxy light indicated by the subtraction. The expected differential slit loss estimated following Section~\ref{sec:dar} ({\galasl}\% at 4000 {\AA}) also seems close to what the best fit in the host-galaxy subtraction indicates (compare the curves at 4000 {\AA} in the right panel).}
\label{fig:multfit}
\end{figure*}

\begin{figure*}
	\centering
        \includegraphics[width=8cm]{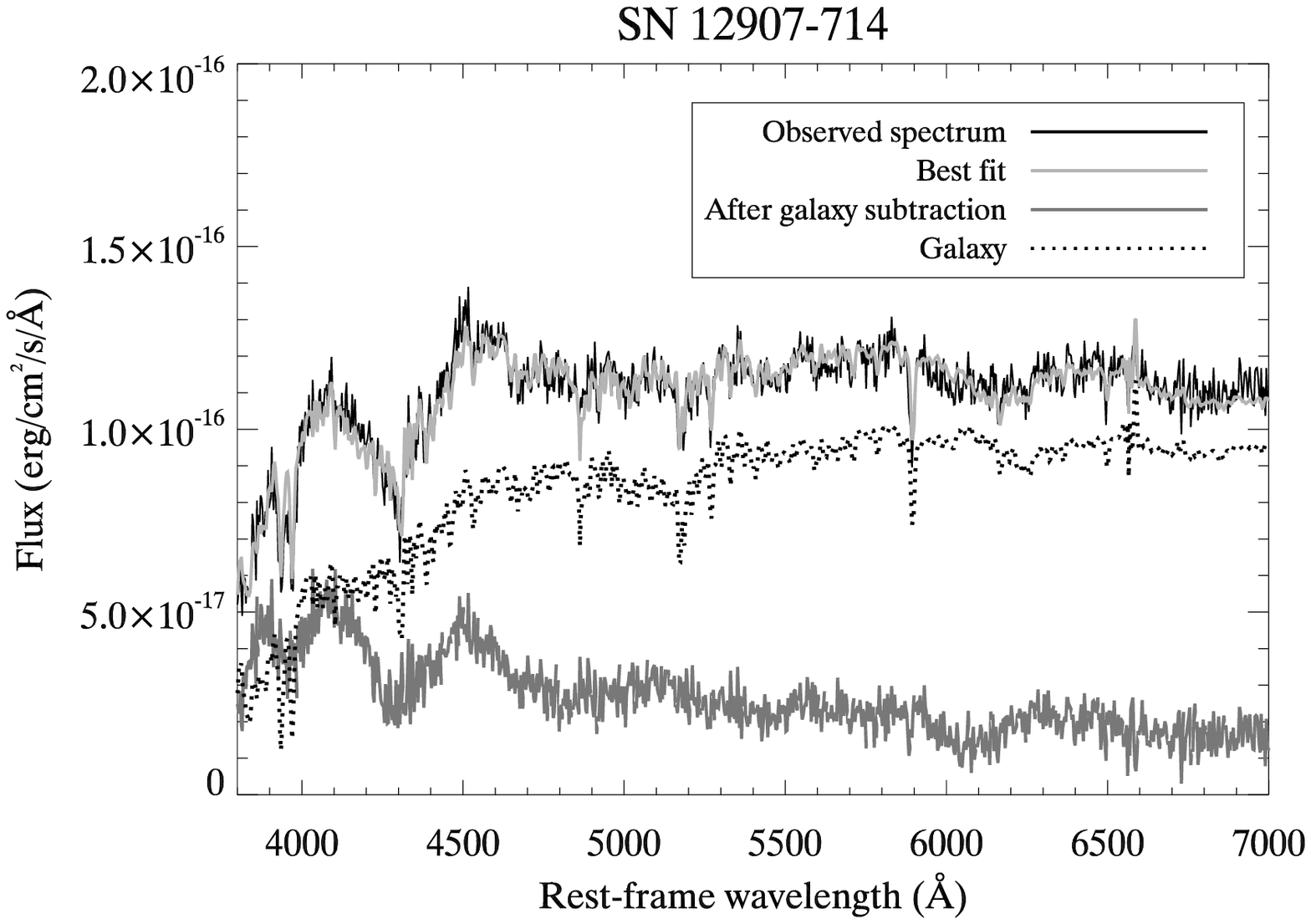}
        \includegraphics[width=8cm]{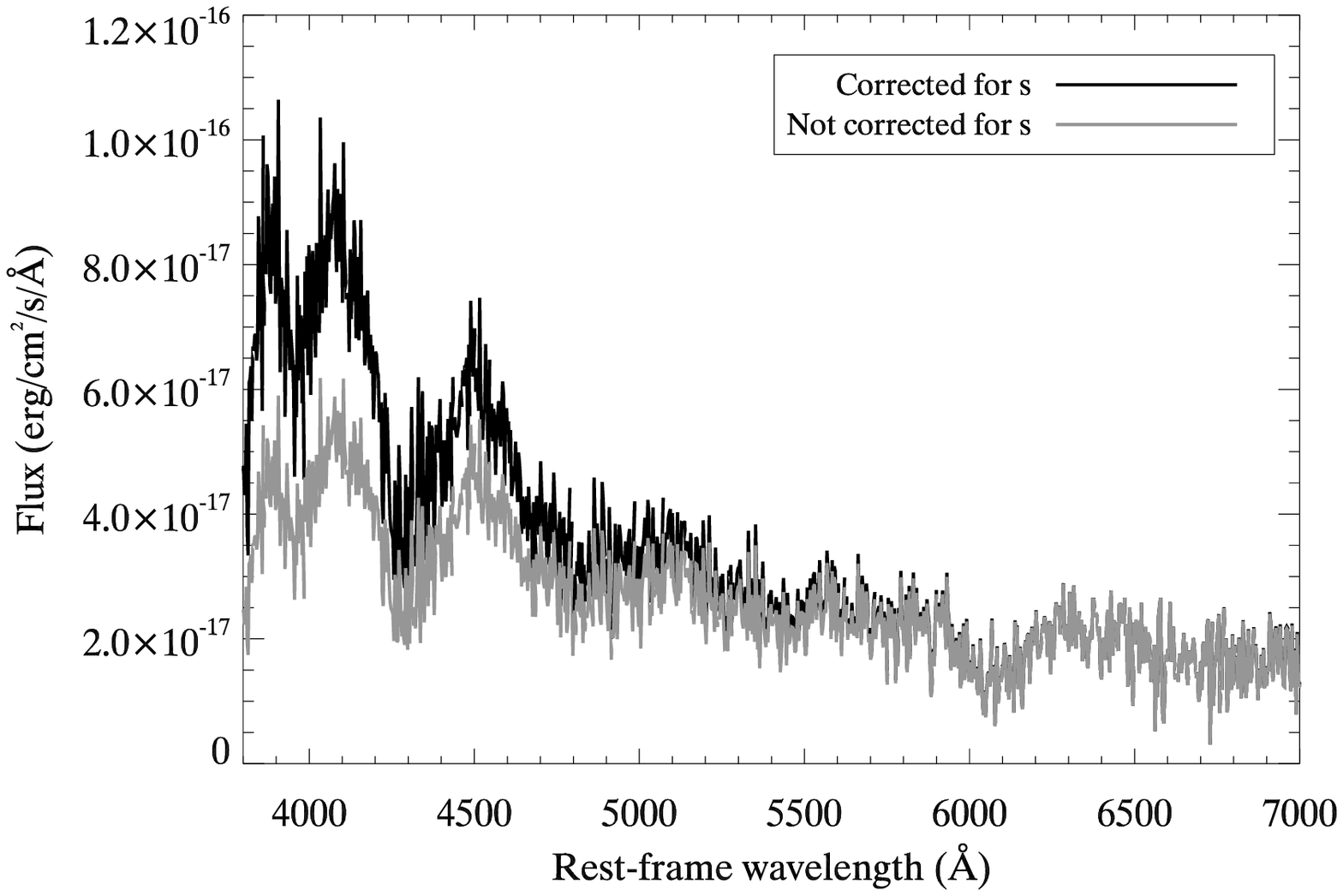}
        \caption{The PCA-based host-galaxy subtraction for the highly contaminated {\snid{12907}} (SN 2006fv) at $z=$ {\galbz}. Left panel: The observed spectrum (black) is plotted together with the best fit (light grey). The galaxy template of the best fit is shown as a dotted line. The grey line shows the SN spectrum after subtraction of the estimated galaxy light. Right panel: The (host subtracted) SN spectrum is shown with and without correction for the wavelength dependent flux loss described by $s$ (black/grey). The host galaxy contamination estimated from photometry ({\galbcont}\%) agrees well with the fraction of host galaxy light indicated by the subtraction. The expected differential slit loss, see Section~\ref{sec:dar}, ({\galbsl}\% at 4000 {\AA}) also seems compatible to what the best fit in the host-galaxy subtraction indicates (compare the curves at 4000 {\AA} in the right panel).}
\label{fig:multfitII}
\end{figure*}

The code for the PCA-based host-galaxy subtraction has only been applied to SNe~Ia with known redshifts and a good lightcurve (photometry both before and after peak brightness).
Two sample subtractions for one moderately and one highly host-galaxy contaminated spectrum can be seen in Figures~\ref{fig:multfit} and~\ref{fig:multfitII}.

\subsection{Host-galaxy simulations}
\label{sec:galeval}

\paragraph{Set-up}

To evaluate the subtraction methods, a large number of simulated 
``\emph{fake}'' spectra were constructed and run through the subtraction pipelines.
Each \emph{fake} spectrum was constructed through a combination of a
SN and a galaxy spectrum, redshifted to some distance. Reddening and
differential slit loss were added to the SN
spectrum. Finally, noise was added. 
All parameters (redshift, spectral epoch, host-galaxy contamination, reddening, differential slit loss) were drawn from distributions that match the NTT/NOT {\dataset}. In
constructing these fake spectra we avoided all templates/functions that
were used in the host-galaxy subtraction pipeline (as this would make the fit
trivial).
More details about the different components of the fake spectra are given below:
\begin{itemize}
\item {\bf{Supernova spectra}} \\
All SN spectra that were used to construct fake spectra have high S/N and low host-galaxy contamination. Their epochs are similar to the ones of the NTT/NOT spectra. Seven different spectra of normal SNe~Ia were used: six of SN 2003du (epochs $-$6, $-$2, 4, 9, 10, 17) \citep{2007AA...469..645S} and one of SN 1998aq at peak brightness \citep{2003AJ....126.1489B}. Furthermore, two spectra of the sub-luminous SN~1999by (epochs $-$5 and 3) \citep{2004ApJ...613.1120G} were used and two of the peculiar and luminous SN 1999aa (epochs $-$5 and 0) \citep{2004AJ....128..387G}.


\item {\bf{Reddening}}\\ 
Reddening was added to the SN spectrum using the
\citet{cardelli89} extinction law with a total-to-selective
extinction ratio $R_V$ of 2.1 and a colour excess $E(B-V)$ drawn from
the distribution of $E(B-V)$ obtained from the NTT/NOT lightcurve
fits.
The lower value of $R_V$ compared to the Milky Way average was chosen to agree with the values often derived when looking at SNe Ia \citep[see e.g.][]{2004MNRAS.349.1344A,2008A&A...487...19N}.

\item {\bf{Galaxy spectra}} \\
In a first simulation series, four galaxy templates of varying type (elliptical, S0, Sa and Sb) from \citet{1996ApJ...467...38K} were used together with three real galaxy spectra (host galaxy spectra for {\snid{7527}}, 13840 and 15381) observed at the NTT at the same time as the SN spectra presented here. The contamination level was randomly chosen between 0 and 70\% for the $g$ band. These simulations were later extended in a second series, where 50 randomly chosen SDSS galaxy spectra were used. Figures displayed here are based on the first run, but results are very similar when including the second set of galaxy spectra.

\item {\bf{Redshifts}} \\
The object redshift was randomly drawn from the NTT/NOT redshift distribution.

\item {\bf{Differential slit loss}}\\
To the SN spectra we multiplied functions modelling the expected differential slit loss due to atmospheric refraction for typical NTT/NOT observing conditions (see Section~\ref{sec:dar}). These effects range from insignificant to severe. The slit loss functions are applied to the SN spectra but not to the galaxy spectra.

\item {\bf{Noise}}\\
An S/N value was randomly chosen from the NTT/NOT spectral S/N distribution. Poisson noise was then added to the spectrum until the target S/N was achieved. The shape of the Poisson noise was determined as a linear combination of the input spectrum and a randomly chosen NTT/NOT sky spectrum. The linear combination was determined such that the highest S/N value in the NTT/NOT sample corresponded to no contribution from sky noise, the lowest S/N corresponded to complete dominance by sky noise and intermediate values to a combination of the two error sources. Simulations were done both with and without noise. As can be expected, added noise increased the dispersion around the true solution, but no significant bias effects were detected.

\end{itemize}

For the simulations without noise, 5000 fake spectra were constructed and about twice as many for the simulations with noise.

\paragraph{Evaluation of the host-galaxy simulations}

\begin{figure*}
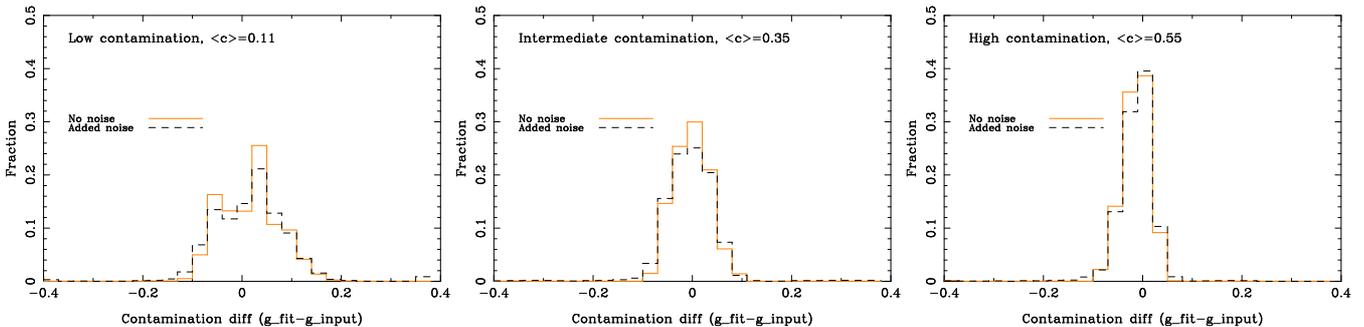

	\centering
      \includegraphics[angle=-90,width=0.65\columnwidth]{15704fg11a.eps}
       \includegraphics[angle=-90,width=0.65\columnwidth]{15704fg11b.eps}
       \includegraphics[angle=-90,width=0.65\columnwidth]{15704fg11c.eps}
      \caption{The difference between the $g$-band host-galaxy contamination that was introduced in the construction of fake spectra ($g\_{\rm{input}}$) and the contamination that was obtained during host-galaxy subtraction ($g\_{\rm{fit}}$) for three different bins in $g\_{\rm{input}}$. The mean contamination for each bin is given in the respective panel as $<c>$. The contamination was calculated as the percentage of the total flux in the observed SDSS $g$ band which comes from the galaxy. The solid histogram corresponds to simulations without added noise while the dashed line corresponds to where noise is added. The dispersion is largest for small input contaminations. The number of outliers increase from 0.9\% to 2\% when noise is added.}
	\label{fig:gcont}
\end{figure*}

The simulations can be evaluated using a range of different tests. For all fake spectra we compared the input SN and galaxy spectra with the SN and galaxy as estimated by the host-galaxy subtraction pipeline. In \citet{nordin10} we study how spectral indicators are affected by the host-galaxy subtraction as a function of contamination level. A more direct way of evaluating subtractions is to compare the input contamination with what is obtained from the subtraction output.
In Figure~\ref{fig:gcont} we show how the \emph{difference} between input and fitted output contamination varies with input contamination (in the $g$ band). The dispersion is largest for low contamination levels. When studying the contribution from different SN templates, it is found that for contamination levels roughly below 20\%, some SN templates yield a small offset. This offset has its origin in a combination of two effects: First, some of the SN templates most likely contain some host galaxy light. This would be seen as a slight bias for low contamination levels, but should thus not be interpreted as a bias of the host subtraction method. Second, a bias can occur if the SN spectrum is imperfectly matched to the Hsiao template. 
The offset decreases with contamination level. For objects with very low contamination it can thus be advantageous not to subtract the estimated host galaxy light since the correction likely is smaller than the uncertainty in the correction.
No trend with galaxy template or slit loss function could be seen.
When noise is added to the fake spectra, the number of failed subtractions increase, from 0.9\%  to 2\%. These failures are usually evident through a visual inspection of the subtraction. The percentage agrees with what is seen when evaluating subtractions of real SNe. These events likely correspond to random noise mimicking host galaxy or SN features.

\subsection{Alternative methods}

Several other host-galaxy subtraction methods were tried: varying the galaxy eigenspectra (their number and source), varying the limitations of the eigenspectra, using a large set of SN templates and using SDSS photometry to fix the fraction of galaxy light as well as the proportions of different galaxy eigenspectra. These alternative methods were found to be either less stable than the method above or equivalent (but involving more computational effort). It should be noted that for some individual SNe~Ia, an alternative method might perform better. Alternative host-galaxy subtractions of spectra can be provided upon request.


\subsection{Discussion on the host-galaxy subtraction}
\label{sec:dischostgal}

The host-subtraction method is not successful for all NTT/NOT SNe, especially for very faint or contaminated cases. It is, however, not obvious how to judge when a subtraction does fail. Both visual inspection and $\chi^2$ tests only indicate how well the subtracted spectrum match the local template, and can thus not be trusted to completely specify which subtractions succeed.

Some of the main worries with the subtraction technique used here are: (i) the observed SNe could differ from the local SN templates used e.g. if it is a peculiar type; (ii) the true galaxy contribution to the spectrum could be badly described by the use of galaxy eigencomponent spectra since the complete galaxy is not recorded in the slit; and (iii) the approximation of slit loss and reddening with a polynomial (applied to the SN but not to the galaxy SED) is a simplification. 
Regarding the first point, object typing was also performed prior to galaxy subtraction, which means that any truly odd SN would have been spotted at this stage unless it is heavily contaminated by galaxy light. Furthermore, the SN template is only used in the choice of what galaxy spectrum to remove and the number of bins used in the fit is large compared to the number of parameters which are fitted and thus the risk of affecting the spectrum severely is small. 
Regarding the second point, the three most dominant eigenspectra have been shown to be able to describe 99\% of all galaxies \citep{2004AJ....128..585Y}, which means that they should be sufficient to describe our host galaxies as well.
The third point, as well as the host-galaxy subtraction code in general, was tested through an extensive number of simulations as described in Section~\ref{sec:galeval}.
The host-galaxy subtraction does fail in roughly one percent of all runs, but these cases usually fail completely, yielding unphysical results, and are thus easily detected.

In \citet{nordin10} we further test the host-galaxy subtraction code, by examining the effects on spectral features, in particular on pseudo-equivalent widths. We also compare the trends of pseudo-equivalent widths for the full sample with the subsample of spectra with low host-galaxy contamination and conclude that no bias is introduced.

%

\section{Typing}
\label{sec:typing}

The object spectra have been typed using SNID
\citep[SuperNova IDentification;][]{2007ApJ...666.1024B}, version
5.0. This algorithm cross-correlates the unknown spectrum against a set of
template spectra. The template database includes SN spectra of
different types and ages as well as non-SN spectra of galaxies,
AGNs and stars. The redshift of the object can either be varied or
fixed. In this version of the code, SNID distinguishes between the following
types and subtypes (within parenthesis): Ia (Ia-norm, Ia-91T, Ia-91bg,
Ia-csm, Ia-pec), Ib (Ib-norm, Ib-pec, IIb), Ic (Ic-norm, Ic-pec,
Ic-broad), II (IIP, II-pec, IIn, IIL), Not-SN (AGN, Gal, LBV,
M-star). The Ia-pec group consists of SNe of the type SN 2000cx
and SN 2002cx.
As a figure of merit for the classification we have used the rlap parameter (see
\citet{2007ApJ...666.1024B} for a definition). A good correlation was considered to be
obtained if rlap was greater or equal to five.
Following \citet{2009AJ....137.3731F} we made four runs with SNID to determine the type, subtype, redshift and age, one
at a time. This was done in the following manner:
\begin{itemize}

\item First we attempted to determine the \emph{type}. A type was
identified if more than 50\% of the templates with a good correlation
belonged to this type and the best-match SN-template (highest
rlap value) had the same type. When the redshift of the
SN was known, we restricted the redshift range of SNID to this
redshift ($\pm 0.02$).

\item If it was possible to determine the type, we continued with trying
to determine the \emph{subtype}. We fixed the type and the redshift
(if it was known). A subtype was considered to be identified if more
than 50\% of the templates with a good correlation belonged to a
specific subtype and the best-match spectrum was of the same
subtype.

\item Regardless if a subtype was identified or not, we determined the
best-fit \emph{redshift}. We now locked SNID to the templates of the
identified type or subtype and fitted for the redshift. The SNID redshift was defined as the median
of the redshifts belonging to templates with a good correlation. The
error was given as the standard deviation.

\item To determine the \emph{age}, the type was fixed to the identified
type or subtype. If we had a spectroscopic redshift, we fixed the
redshift in SNID to this value. Otherwise we used the value obtained in
the step before with a redshift error of 0.02. The SNID age was then
defined as the median of all ages belonging to templates with a good
correlation. The error of the age was given as the standard deviation
of these values.

\end{itemize}

If the determined type and redshift were inconsistent in that the preferred type was an M-star or an LBV, while the redshift was too high for such an object to be detected, the type was set to unknown.

%
A problem with this method is that the composition of the
template database affects the possibilities to detect different
types. The SNID database has few spectra of core-collapse SNe
and peculiar SNe~Ia which makes these objects harder to type. It is also less
effective for low S/N spectra than for those with high S/N.

%
The outcome of the SNID analysis is presented in Table~\ref{tab:type}. The typing for each
spectrum is presented in the column \emph{NTT/NOT type}. When the
type was determined using the host-galaxy subtracted spectrum, this is
marked in the column for notes with the addition of an $s$.
In some cases a spectrum only had a good match using SNID with one particular object in the database. In these cases we have marked this in the table. 
When no type could be determined this is marked with ``$-$''.

%
The consistency in typing when several spectra of the same
object existed was good. If a
type could be determined, the same type was always obtained. In some cases, one of the spectra was left
without a type due to, for example, noise making the classification difficult or it was classified as a galaxy, due to the faintness of the SN at that epoch.

%
After a visual inspection of the spectra, the typing from SNID was changed for {\nohandtyped} spectra. 
The majority of these, {\nofromunknown}, could not be typed by SNID but could be clearly typed by visual inspection. 
There was also a group of spectra, {\notounknown}, which had been typed by SNID, but with only few well matched template spectra, and, in the visual study, they were found to be of such low quality that it was not possible to say with certainty what the type was. {\nochangetext} spectra had a type which was changed, one from a Ib to a Ib/c and two from a Ia to a possible Ia. There were also {\noremsubtext} cases where the type was kept, but the subtype was removed.
The spectra whose type was changed following the visual check are marked with a $v$ in the column for notes in the table.

%
The NTT/NOT objects also have an SDSS typing based on the NTT and the NOT spectra in combination with
spectra taken at other telescopes as a part of the SDSS-II SN Survey. This type is presented in Table~\ref{tab:type} under the column \emph{SDSS type}. Due to the faintness of the SN in some spectra, not all NTT and NOT spectra could be
typed and some were typed as galaxies, while the overall SDSS typing
is a SN, based on spectra at other epochs from other
telescopes. 

\begin{figure}
	\centering
		\includegraphics[width=8cm]{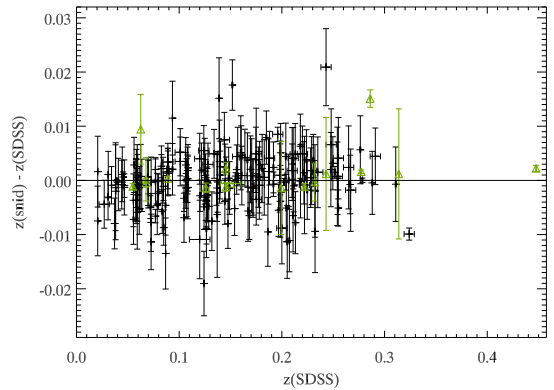}
	\caption{A comparison of the SDSS object redshifts with the redshifts determined using
	SNID. The vertical error bars show the uncertainty in the SNID
	redshifts. The horizontal error bars show the uncertainty in the SDSS object redshifts. In most cases these are too small to be visible. All redshifts determined from SN features have an error of at least 0.005. The RMS of the distribution is {\zdisp} (with a negligible
        bias). SNID redshifts which are calculated using
	less than five well matched spectra are marked with green triangles.}
	\label{fig:snid_z}
\end{figure}

In Figure~\ref{fig:snid_z} the redshifts obtained from SNID are compared with the SDSS object redshifts from Zheng et al. (in preparation), which mainly come from the SDSS DR7 catalogue and measurements of galaxy lines in spectra obtained as a part of SDSS-II. Some are determined from SN features (see Section~\ref{sec:data}).
There is a good agreement between the SNID redshifts and the SDSS redshifts, with a dispersion of {\zdisp} and a negligible
bias. No dependence with redshift is detected.

\begin{figure}
	\centering
		\includegraphics[width=8cm]{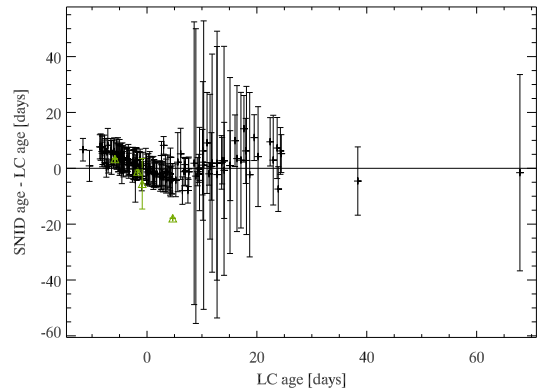}
	\caption{A comparison of the age obtained using SNID and the
	age obtained from the lightcurve fit using SALT. The age is
	here defined as the number of days in rest frame since $B$-band
	maximum. SNID ages which are calculated using less than five well matched spectra are marked with green triangles.}
	\label{fig:snid_age}
\end{figure}

In Figure~\ref{fig:snid_age}, the ages of the spectra as estimated with SNID are
compared with the ages obtained from the lightcurves, $({\rm{MJD}}_{\rm{spec}}-{\rm{MJD}}_{\rm{max}})/(1+z)$, for the spectra which we have classified
as SNe~Ia.
A dispersion of {\agedisp} days was obtained. It should be noted that for young
SNe, SNID generally estimates an older age compared to the SALT
lightcurve fit and that the errors in the age
from SNID are typically overestimated, especially at later epochs. This
was also pointed out by \citet{2007ApJ...666.1024B}.
Since the estimated age error bars are symmetric (estimated as the standard deviation) and in many cases overestimed, there will be some cases where the minimum allowed age (from the error bar) will be unphyically low.
We have \emph{not} used any prior on the age
from the lightcurve, which cause the large error bars and the bias in
the epoch estimates at early epochs.
It should be noted that the error bars, in a few cases, also could be \emph{underestimated}. This occurs when the observed spectrum is only well fitted with few template SNe and these template epochs do not sample all different ages of a SN. One example is {\snid{16838}} which is only well fit by spectra in the library of SN 1998S. For {\noonesnidagetext} spectrum ({\snid{\noonesnidageid}}), no error was estimated since only one template spectrum could fit the observed spectrum.
We also test the hypothesis that the evolution of spectral features scale with the stretch factor, by multiplying the age obtained from SNID with $s$. The bias is slightly improved, while the spread increases.

%

\section{Some special objects}
\label{sec:special}

\paragraph{Possible non-normal SNe Ia}

\begin{figure}
	\centering
	\includegraphics[width=9cm]{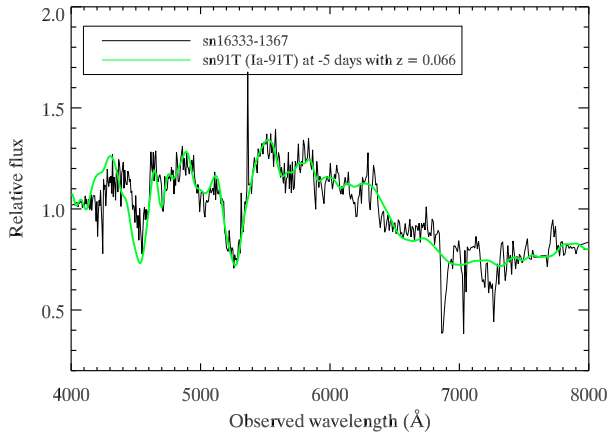}
	\caption{SNID fit for {\snid{16333}} (SN 2006on) with a SN~1991T template at 5 days before lightcurve maximum.}
	\label{fig:sn16333}
\end{figure}

Using SNID we find that the spectrum of {\snid{16333}} (\object{SN 2006on}) is only well described by spectra of SN~1991T. The fit with SN 1991T at 5 days before maximum brightness is shown in Figure~\ref{fig:sn16333}.
The peak brightness of the SN is well constrained by the lightcurve fit (the stretch and colour less so) and we find a peak $B$ absolute magnitude of $-$19.06. The obtained stretch was 0.94 and the SALT $c$ colour 0.19. Correcting the peak magnitude for stretch and colour, following 
$M_B + \alpha (s-1) - \beta c$, gives a peak magnitude of $-$19.6, which is consistent with a SN 1991T-like SN.

\begin{figure}
	\centering
	\includegraphics[width=9cm]{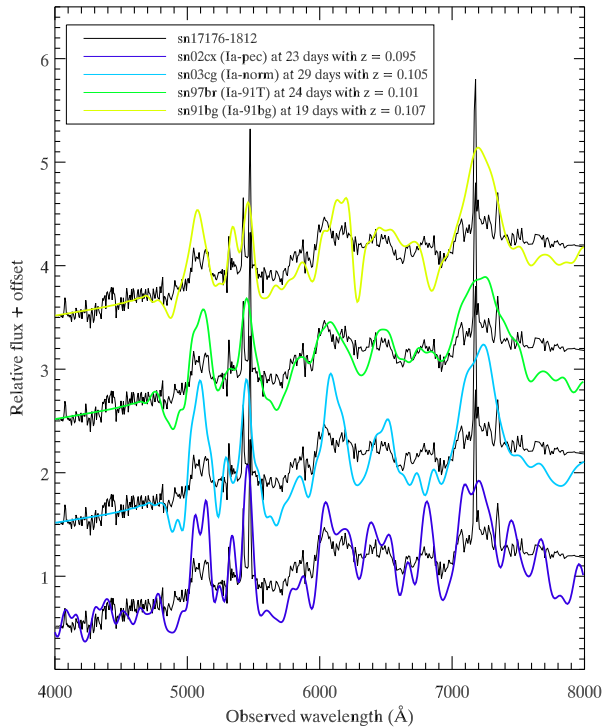} 
	\caption{The observed spectrum of {\snid{17176}} (SN 2007ie) in black together with different template spectra as indicated in the legend. In the fit, the redshift has been fixed to 0.0935, the redshift of the object as obtained from SDSS DR7, with an allowed variation of 0.02. The fit with the spectrum of SN 2002cx gives the best match and the closest redshift compared to the one of the host galaxy.}
	\label{fig:sn17176}
\end{figure}

The spectrum of {\snid{17176}} (\object{SN 2007ie}) has strong indications of being a SN 2002cx-like object.
Using SNID, the spectrum is best fit with a spectrum of SN 2002cx. However, also other classes of SNe Ia give sufficiently good fits to the spectrum, when allowing for a variation of 0.02 in redshift. In Figure~\ref{fig:sn17176}, some of the SNID fits with different template spectra are shown.
The spectrum of SN 2002cx is a better fit to the double peaks at an observed wavelength of 5100 {\AA} and 6100 {\AA} than the other spectra (normal SN~Ia, SN~1991T and SN~1991bg). Furthermore, the fit with the spectrum of SN 2002cx gives a better match with the redshift of the SN as obtained from SDSS DR7. If we reduce the allowed redshift range in SNID to $\pm 0.01$, SNID identifies the spectrum as a SN 2002cx-like object.
By comparing the spectrum with the ones by \citet{2008ApJ...680..580S} of SN 2005hk, which also is a SN 2002cx-like SN, we identify absorption by Fe~{\sc{ii}}~$\lambda$5018, Fe~{\sc{ii}}~$\lambda$5535, Fe~{\sc{ii}}~$\lambda$6149 and Fe~{\sc{ii}}~$\lambda$6247. The velocities in our spectrum for these lines are lower by 7$-$35\% compared to a spectrum of SN 2005hk at 24 days past peak brightness.
Compared to the spectrum at 38 days past peak, the velocities varies from being lower by 26\% to being higher by 9\%.
The lightcurve, in this case, is not well constrained due to lack of photometry before maximum brightness. The best-fit absolute peak magnitude in the $B$-band from SALT is
-18.2, which is comparable to what is expected for SN 2002cx-like objects. 

\begin{figure}
	\centering
	\includegraphics[width=9cm]{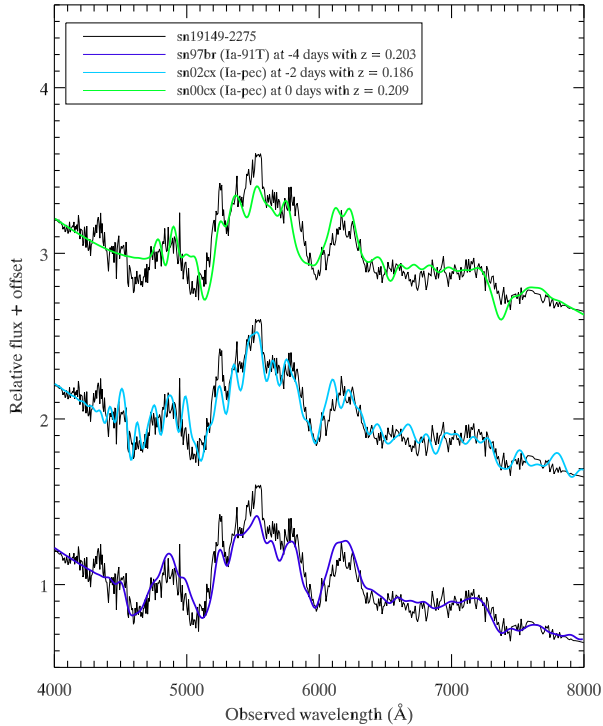}
	\caption{The SNID fit of {\snid{19149}} (SN 2007ni) with the observed spectrum in black and different template spectra in colour as indicated in the legend.}
	\label{fig:sn19149}
\end{figure}

{\snid{19149}} (\object{SN 2007ni}) is best fit with a SN 1991T-like spectrum according to SNID. It also has great similarities with SN 2002cx-like and SN 2000cx-like spectra, but not with spectra of normal SNe Ia. Comparisons from SNID with different templates are shown in Figure~\ref{fig:sn19149}. 
The lightcurve fit is well constrained and the obtained absolute peak magnitude in $B$ is $-$19.47, which is somewhat brighter than a normal SN Ia. We could thus exclude an SN~2002cx interpretation. The stretch and SALT $c$ colour are 1.04 and 0.09, respectively. 


{\snid{20978} (\object{SN 2007rl}) is best fit in SNID with a spectrum of SN 2000cx. However, due to its high redshift, it is well fitted by many different templates.
The peak brightness is well constrained by the photometry, while the stretch and colour are less so due to few observations past peak brightness. The absolute peak $B$ band magnitude is $-$19.6.}

\paragraph{Possible SNe~IIn}

In the spectral fitting we find that the spectrum of {\snid{16838}} is best fit with a SN~IIn, similar to SN 1998S. Another potential SN~IIn is {\snid{16668}} (\object{SN 2006pu}).

\paragraph{Stellar tidal disruption event}

The object with SDSS identification 17237 has been suggested to be a stellar tidal disruption event 
(van Velzen et al., in preparation), an event where a star passes close to a supermassive black hole in the centre of the
galaxy without being completely destroyed.


\section{Summary}
\label{sec:results}

We have presented observations and reductions of
{\nospectra} spectra observed at the NTT and the NOT as a part of a
programme to classify SNe discovered by the SDSS-II. The selection procedure, the
observing strategy and the reductions have been performed in a coherent and well documented manner. 
The spectra have been corrected for telluric absorption. Objects classified as SNe Ia were processed through a host-galaxy subtraction pipeline. 
The host galaxy SEDs were estimated using a PCA-analysis where the difference between the observed spectrum and a combination of a SN template and galaxy eigencomponent spectra were minimised. A polynomial was multiplied to the SN templates to account for reddening effects as well as differential slit loss due to atmospheric refraction. The host-galaxy subtraction pipeline was evaluated using extensive simulations
mimicking the distribution of parameters for the NTT/NOT {\dataset}.
Furthermore, careful estimates of uncertainties have been
made, thus making our data well-suited for systematic studies.

The spectra were classified using SNID, a template-matching algorithm. The age of the spectra estimated using SNID was compared to the age obtained from the lightcurve, in the case of SNe Ia, and was found to agree well, with an age dependent bias of a few days and an overall dispersion of {\agedisp} days. The typing was revised using visual inspection and changed in some cases; typically when the SNID classification was unreliable due to few matched spectra or when the spectrum could be typed by visual inspection but not with SNID.

In total, {\nosnspectra} spectra of {\nosne} different confirmed SNe were
obtained at the NTT and the NOT for the SDSS-II SN program. Out
of these, {\nosniaspectra} are SN~Ia spectra, and some of these (the ones which will survive the strict cosmology lightcurve cuts) will 
be part of a larger sample to be included in the three year SDSS-II SN
Hubble diagram. 
Four potential peculiar SNe Ia were identified from their spectra. {\snid{16333}} (SN 2006on) and 19149 (SN 2007ni) are potential SN 1991T-like objects, while {\snid{17176}} (SN 2007ie) is a potential SN 2002cx object.  {\snid{20978}} (SN 2007rl) is identified as a peculiar SN Ia using SNID, but due to its high redshift, it is difficult to determine a subtype.
The sample also includes {\nosnIIspectra} spectra of {\nosneII} SNe~II 
and {\nosnIbcspectra} spectra of {\nosneIbc} SNe~Ib/c.

The reduced spectra are made public together with the estimated
uncertainty for each wavelength. For the subsample of SNe Ia, host-galaxy subtracted spectra are provided. 
All this data is accessible from the webpage:
\texttt{http://www.physto.se/}$\sim$\texttt{linda/spectra/nttnot.html}


\begin{acknowledgements}

The authors would like to thank Johan Fynbo, Christa Gall and
Christina Henriksson who all helped out at the NOT observations.

L.{\"{O}} is partially supported by the Spanish Ministry of Science and Innovation (MICINN) through the Consolider Ingenio-2010 program, under project CSD2007-00060 ``Physics of the Accelerating Universe (PAU)''.

V.S is financially supported by FCT Portugal under program Ci\^{e}ncia 2008.

The Oskar Klein Centre is funded by the Swedish Research Council.
The Dark Cosmology Centre is funded by the Danish National Research
Foundation.

Funding for the SDSS and SDSS-II has been provided by the Alfred P. Sloan Foundation, the Participating Institutions, the National Science Foundation, the U.S. Department of Energy, the National Aeronautics and Space Administration, the Japanese Monbukagakusho, the Max Planck Society, and the Higher Education Funding Council for England. The SDSS Web Site is http://www.sdss.org/.

The SDSS is managed by the Astrophysical Research Consortium for the Participating Institutions. The Participating Institutions are the American Museum of Natural History, Astrophysical Institute Potsdam, University of Basel, University of Cambridge, Case Western Reserve University, University of Chicago, Drexel University, Fermilab, the Institute for Advanced Study, the Japan Participation Group, Johns Hopkins University, the Joint Institute for Nuclear Astrophysics, the Kavli Institute for Particle Astrophysics and Cosmology, the Korean Scientist Group, the Chinese Academy of Sciences (LAMOST), Los Alamos National Laboratory, the Max-Planck-Institute for Astronomy (MPIA), the Max-Planck-Institute for Astrophysics (MPA), New Mexico State University, Ohio State University, University of Pittsburgh, University of Portsmouth, Princeton University, the United States Naval Observatory, and the University of Washington.

The paper is partly based on observations made with the Nordic Optical
Telescope, operated on the island of La Palma jointly by Denmark,
Finland, Iceland, Norway, and Sweden, in the Spanish Observatorio del
Roque de los Muchachos of the Instituto de Astrofisica de Canarias.
The data have been taken using ALFOSC, which is owned by the Instituto
de Astrofisica de Andalucia (IAA) and operated at the Nordic Optical
Telescope under agreement between IAA and the NBI.

The paper is partly based on observations collected at the New Technology Telescope, operated by the European Organisation for Astronomical Research in the Southern Hemisphere, Chile.

\end{acknowledgements}


\bibliographystyle{aa}
\bibliography{15704bib}


\listofobjects


\appendix

\section{Tables}
\label{sec:tables}

\longtab{1}{
\begin{longtable}{p{1.0cm}p{0.9cm}p{1.2cm}p{1.4cm}p{1.6cm}p{1.4cm}p{1.4cm}p{1.4cm}}
\caption{\label{tab:obs} Information about the observations.} \\
\hline \hline
ID & SPID & IAU \tablefootmark{a} & Telescope & Slit Width (arcsec) & Exposure Time (s) & Airmass &  Seeing (arcsec) \tablefootmark{b}\\
\hline
\endfirsthead
\caption{continued.}\\
\hline\hline
ID & SPID & IAU \tablefootmark{a} & Telescope & Slit Width (arcsec) & Exposure Time (s) & Airmass &  Seeing (arcsec) \tablefootmark{b}\\
\hline
\endhead
\hline
\multicolumn{8}{l}{{Continued on next page\ldots}} \\
\endfoot
\hline
\endlastfoot
12778&692&2006fs&NTT&1.0&    1000&1.16&1.1\\
12779&693&2006fd&NTT&1.5&    1000&1.27&1.2\\
12781&680&2006er&NTT&1.5&    1800&1.14&1.9\\
12782&681&2006fq&NTT&1.5&    1000&1.27&1.6\\
12820&711&2006fg&NTT&1.5&     600&1.41&1.8\\
12842&682&2006ez&NTT&1.5&     800&1.14&1.5\\
12843&727&2006fa&NTT&1.0&    1000&1.15&0.8\\
12844&684&2006fe&NTT&1.0&    2000&1.21&1.6\\
12853&685&2006ey&NTT&1.5&    1000&1.17&1.1\\
12855&716&2006fk&NTT&1.5&    2000&1.25&2.0\\
12856&695&2006fl&NTT&1.5&    2197&1.38&1.5\\
12860&688&2006fc&NTT&1.5&    1000&1.16&1.6\\
12874&689&2006fb&NTT&1.5&    2400&1.20&2.5\\
12898&712&2006fw&NTT&1.5&     600&1.35&1.5\\
12907&714&2006fv&NTT&1.5&     600&1.21&1.6\\
12927&690&2006fj&NTT&1.5&    1800&1.25&2.5\\
12928&686&2006ew&NTT&1.0&    2000&1.20&1.2\\
12930&687&2006ex&NTT&1.5&    1000&1.16&1.4\\
12947&691&&NTT&1.5&    1500&1.36&1.7\\
12950&700&2006fy&NTT&1.0&     600&1.19&1.1\\
12950&1055&2006fy&NTT&1.0&     600&1.37&1.1\\
12978&701&&NTT&1.5&    1800&1.59&3.0\\
13005&702&2006fh&NTT&1.5&    2000&1.14&2.0\\
13025&761&2006fx&NTT&1.0&    1200&1.24&0.8\\
13044&724&2006fm&NTT&1.0&     900&1.19&1.0\\
13044&1062&2006fm&NTT&1.0&    1200&1.60&1.2\\
13045&734&2006fn&NTT&1.0&    1000&1.41&1.7\\
13046&726&&NTT&1.0&    1500&1.46&1.4\\
13070&736&2006fu&NTT&1.0&    2000&1.19&0.9\\
13072&723&2006fi&NTT&1.0&    1500&1.30&1.7\\
13135&739&2006fz&NTT&1.0&    1200&1.19&1.7\\
13135&998&2006fz&NTT&1.0&     600&1.26&1.2\\
13174&766&2006ga&NTT&1.5&    1000&1.48&1.6\\
13195&764&2006fo&NTT&1.0&     300&2.04&1.7\\
13195&983&2006fo&NTT&1.0&     900&1.66&1.7\\
13195&1458&2006fo&NTT&1.0&     900&2.00&0.8\\
13355&1003&2006kh&NTT&1.0&     900&1.47&1.4\\
13376&1002&2006gq&NTT&1.0&    1200&1.36&1.4\\
13376&1106&2006gq&NTT&1.0&    1800&1.66&1.4\\
13577&1000&2006kg&NTT&1.0&     900&1.62&1.7\\
13796&1058&2006hl&NTT&1.0&     900&1.51&1.3\\
13894&1039&2006jh&NTT&1.0&    1800&1.35&1.1\\
14157&1040&2006kj&NTT&1.0&    1800&1.26&1.0\\
14279&1459&2006hx&NTT&1.0&     900&1.87&0.9\\
14318&1594&2006py&NTT&1.0&     900&2.47&1.0\\
14318&1653&2006py&NTT&1.0&     600&1.99&1.6\\
14318&1713&2006py&NTT&1.5&    1800&1.77&2.3\\
14437&1061&2006hy&NTT&1.0&     900&1.43&1.2\\
14450&991&2006kn&NTT&1.0&    1800&1.21&1.0\\
14451&989&2006ji&NTT&1.0&    1200&1.42&1.5\\
14492&1001&2006jo&NTT&1.0&    1800&1.31&1.4\\
14598&987&&NTT&1.0&     900&1.20&1.3\\
14599&988&2006jl&NTT&1.0&     600&1.16&1.1\\
14782&990&2006jp&NTT&1.0&     900&1.25&1.8\\
14846&1014&2006jn&NTT&1.0&    1200&1.32&1.2\\
14871&1008&2006jq&NTT&1.0&     600&1.17&1.8\\
14979&1009&2006jr&NTT&1.5&     900&1.21&1.9\\
14984&1027&2006js&NTT&1.0&    1800&1.18&1.1\\
15031&985&2006iw&NTT&1.0&     450&1.17&1.6\\
15129&1015&2006kq&NTT&1.0&    1800&1.24&1.0\\
15132&1012&2006jt&NTT&1.0&     900&1.20&1.8\\
15136&1022&2006ju&NTT&1.0&     900&1.33&1.2\\
15153&1046&&NTT&1.0&     900&1.26&1.7\\
15161&1010&2006jw&NTT&1.0&    1800&1.48&1.3\\
15171&1045&2006kb&NTT&1.0&    1800&1.26&1.4\\
15203&1026&2006jy&NTT&1.0&    1800&1.48&1.8\\
15207&1038&&NTT&1.0&    1200&1.51&1.6\\
15210&1005&&NTT&1.0&    1800&1.23&1.3\\
15210&1052&&NTT&1.0&    1800&1.30&1.1\\
15222&1004&2006jz&NTT&1.0&    1800&1.53&1.3\\
15234&1043&2006kd&NTT&1.0&     900&1.69&2.3\\
15259&1051&2006kc&NTT&1.0&    1800&1.30&1.5\\
15287&1057&2006kt&NTT&1.0&    1800&1.29&1.0\\
15320&1098&2006kv&NTT&1.0&    1200&1.50&1.6\\
15339&1107&2006ns&NTT&1.0&    1800&1.48&1.1\\
15354&1110&2006lp&NTT&1.0&    1800&1.54&1.6\\
15475&1464&2006lc&NTT&1.0&     900&2.26&1.3\\
15557&1532&2006oz&NOT&1.3&    2000&1.30&1.0\\
16021&1355&2006nc&NOT&1.3&    1800&1.14&1.4\\
16069&1358&2006nd&NOT&1.3&    2000&2.67&1.9\\
16069&1467&2006nd&NTT&1.0&    1200&1.67&1.0\\
16069&1651&2006nd&NTT&1.5&    1800&1.78&2.0\\
16087&1455&2006pc&NTT&1.0&    1800&1.67&0.8\\
16163&1678&&NTT&1.5&    1800&1.57&1.9\\
16165&1326&2006nw&NOT&1.3&    2100&1.60&1.5\\
16179&1323&2006nx&NOT&1.3&    1800&2.08&1.4\\
16179&1469&2006nx&NTT&1.0&    1200&1.83&0.6\\
16179&1570&2006nx&NTT&1.0&    1200&2.21&0.6\\
16192&1322&2006ny&NOT&1.3&    1400&1.75&1.3\\
16192&1496&2006ny&NTT&1.0&    1800&1.73&1.6\\
16204&1500&&NTT&1.0&     900&2.02&1.5\\
16206&1501&2006pe&NTT&1.0&    1200&1.51&1.0\\
16215&1456&2006ne&NTT&1.0&     900&2.02&0.9\\
16215&1630&2006ne&NTT&1.0&    1800&1.83&1.4\\
16241&1470&&NTT&1.0&    1800&1.75&1.0\\
16280&1471&2006nz&NTT&1.0&     900&1.50&0.6\\
16280&1564&2006nz&NTT&1.0&    1200&1.84&1.8\\
16287&1449&2006np&NTT&1.0&     900&1.81&0.6\\
16287&1569&2006np&NTT&1.0&     900&2.26&0.7\\
16287&1650&2006np&NTT&1.0&    1800&1.70&1.7\\
16302&1473&&NTT&1.0&    1800&2.10&1.9\\
16314&1335&2006oa&NOT&1.3&    1200&1.31&1.7\\
16314&1475&2006oa&NOT&1.3&    1000&1.18&0.9\\
16333&1367&2006on&NOT&1.3&    2000&1.48&1.3\\
16352&1478&2006pk&NTT&1.0&    1800&1.82&1.1\\
16391&1452&&NTT&1.0&    1200&1.69&0.9\\
16391&1565&&NTT&1.0&    1800&1.81&0.9\\
16392&1365&2006ob&NOT&1.3&    1100&3.06&1.9\\
16392&1448&2006ob&NTT&1.0&     900&1.89&0.8\\
16392&1566&2006ob&NTT&1.0&    1200&2.18&2.3\\
16392&1682&2006ob&NTT&1.5&    1800&1.43&1.8\\
16402&1505&2006sv&NTT&1.0&    1200&2.38&0.9\\
16473&1520&2006pl&NTT&1.0&    1800&1.85&1.0\\
16541&1485&2006pn&NOT&1.3&    2400&1.76&1.4\\
16578&1516&2006po&NTT&1.0&    1800&2.19&1.2\\
16619&1519&2006ps&NTT&1.0&     900&2.05&0.9\\
16619&1528&2006ps&NOT&1.3&    1800&1.30&1.4\\
16637&1514&&NTT&1.0&    1800&1.47&0.7\\
16641&1518&2006pr&NTT&1.0&     900&1.88&0.8\\
16641&1530&2006pr&NOT&1.3&    1800&1.17&1.3\\
16641&1649&2006pr&NTT&1.0&    1200&2.00&1.5\\
16668&1561&2006pu&NOT&1.3&    2000&1.33&1.3\\
16692&1489&2006op&NTT&1.0&     900&1.85&1.0\\
16737&1599&2006qc&NTT&1.0&    1200&2.29&0.7\\
16741&1523&&NOT&1.3&     600&1.23&1.1\\
16748&1574&2006sx&NTT&1.0&    1800&1.95&0.8\\
16774&1606&2006sy&NTT&1.0&    1800&2.04&0.9\\
16778&1542&&NTT&1.0&    1800&1.83&0.6\\
16778&1568&&NTT&1.0&    1800&1.66&0.8\\
16793&1603&2006qg&NTT&1.0&    1200&2.32&0.9\\
16838&1522&&NOT&1.3&    2500&1.45&1.1\\
16857&1538&&NTT&1.0&    1200&1.93&0.6\\
16867&1541&&NTT&1.0&    1800&1.73&1.1\\
16872&1539&2006qh&NTT&1.0&    1200&2.00&0.6\\
16956&1562&2006qj&NTT&1.0&    1500&2.34&0.9\\
16979&1597&&NTT&1.0&    1800&2.07&1.0\\
16988&1595&2006qk&NTT&1.0&    1200&2.27&0.8\\
16988&1652&2006qk&NTT&1.0&    1800&1.99&1.5\\
17117&1679&2006qm&NTT&1.0&    1800&1.96&1.4\\
17135&1648&2006rz&NTT&1.0&    1800&1.92&1.8\\
17167&2250&2007mr&NTT&1.0&    1800&1.29&1.2\\
17170&1879&&NTT&1.0&    1800&1.22&1.4\\
17176&1812&2007ie&NTT&1.5&    1200&1.16&1.7\\
17200&1796&2007ja&NTT&1.5&    1800&1.32&1.7\\
17206&1788&&NTT&1.5&    1800&1.27&1.4\\
17218&1794&2007jp&NOT&1.0&    2000&1.14&0.8\\
17220&1791&2007ji&NTT&1.5&    1200&1.26&1.8\\
17223&1793&2007jj&NTT&1.5&    1800&1.21&2.3\\
17237&1830&2007jc&NTT&1.5&    1200&1.15&2.3\\
17245&2234&&NTT&1.0&    1800&1.20&0.9\\
17247&1799&&NTT&1.0&    1800&1.18&1.1\\
17253&1898&2007jq&NOT&1.0&    2000&1.16&0.8\\
17254&1813&2007ii&NTT&1.5&    1800&1.21&1.8\\
17332&1899&2007jk&NOT&1.0&    1400&1.20&0.9\\
17351&1769&2007jy&NTT&1.0&    1800&1.17&1.1\\
17366&1782&2007hz&NTT&1.0&    1200&1.14&1.4\\
17389&1811&2007ih&NTT&1.0&    1800&1.14&1.6\\
17391&1872&2007jo&NTT&1.5&    1800&1.18&1.5\\
17422&1785&&NTT&1.0&    1800&1.18&0.9\\
17435&1902&2007ka&NOT&1.0&    2000&1.25&1.0\\
17436&1790&&NTT&1.5&    1800&1.34&1.9\\
17464&1853&2007jb&NTT&1.5&    2400&1.47&2.5\\
17486&1854&&NTT&1.0&    1800&1.14&1.9\\
17497&1837&2007jt&NOT&1.0&    2000&1.17&0.8\\
17500&2249&2007lf&NTT&1.0&    1800&1.19&1.0\\
17535&1838&&NOT&1.0&    2000&1.16&0.8\\
17548&1825&2007ms&NTT&1.0&    1200&1.15&1.1\\
17548&2231&2007ms&NTT&1.0&    1200&1.25&0.9\\
17548&2293&2007ms&NTT&1.5&    1800&1.24&2.0\\
17552&1789&2007jl&NTT&1.5&    1800&1.29&1.7\\
17568&1810&2007kb&NTT&1.0&    1800&1.17&1.2\\
17605&1809&2007js&NTT&1.5&    1200&1.33&1.8\\
17627&1841&2007jf&NOT&1.0&    2000&1.16&0.7\\
17629&1851&2007jw&NTT&1.5&    1800&1.29&2.0\\
17647&1875&&NTT&1.0&    1800&1.42&1.4\\
17703&1881&&NTT&1.5&    1800&1.25&1.9\\
17745&2161&2007ju&NTT&1.5&    1800&1.32&2.8\\
17746&1873&2007jv&NTT&1.5&    1800&1.16&1.7\\
17784&1842&2007jg&NOT&1.0&     700&1.19&0.7\\
17790&1887&2007jx&NOT&1.0&    2000&1.28&1.0\\
17794&1906&&NTT&1.0&    1800&1.16&1.1\\
17811&1816&2007ix&NTT&1.5&    1800&1.14&2.2\\
17814&1901&&NTT&1.0&    1800&1.14&0.9\\
17825&1819&2007je&NTT&1.5&    1800&1.25&2.7\\
17854&2230&&NTT&1.0&    1800&1.48&0.8\\
17875&1817&2007jz&NTT&1.0&    1800&1.23&1.3\\
17880&1843&2007jd&NOT&1.0&     900&1.20&0.7\\
17880&1957&2007jd&NTT&1.5&    1800&1.26&2.3\\
17880&2194&2007jd&NTT&1.5&    1800&1.51&2.0\\
17886&1844&2007jh&NOT&1.0&     600&1.13&0.9\\
17924&1826&&NTT&1.5&    1200&1.20&2.2\\
17973&1926&&NOT&1.0&     900&1.14&0.9\\
17973&1942&&NTT&1.0&    1200&1.19&0.9\\
18109&1940&2007kw&NTT&1.0&    1200&1.19&1.3\\
18325&2277&2007mv&NTT&1.0&    1800&1.21&0.9\\
18457&2285&2007ll&NTT&1.5&    1800&1.38&2.4\\
18466&2270&2007lm&NTT&1.0&    1800&1.34&1.0\\
18590&2248&2007nw&NTT&1.0&    1800&1.35&0.8\\
18596&2227&2007ld&NTT&1.5&    1500&1.31&2.1\\
18647&2271&&NTT&1.0&    1800&1.22&0.8\\
18697&2171&2007ma&NTT&1.5&    1800&1.33&2.6\\
18768&2135&2007lh&NTT&1.0&    1800&1.24&1.0\\
18787&2150&2007mf&NTT&1.0&    1800&1.34&0.8\\
18804&2148&2007me&NTT&1.0&    1500&1.22&1.0\\
18903&2247&2007lr&NTT&1.0&    1800&1.18&1.0\\
18965&2279&2007ne&NTT&1.0&    1800&1.29&0.9\\
19003&2235&2007mp&NTT&1.0&    1800&1.42&0.8\\
19003&2290&2007mp&NTT&1.0&    1800&1.35&1.9\\
19008&2284&2007mz&NTT&1.0&    1800&1.15&0.8\\
19023&2236&2007ls&NTT&1.0&    2000&1.21&1.1\\
19051&2297&2007nb&NTT&1.0&    1800&1.18&0.8\\
19101&2268&2007ml&NTT&1.0&     543&1.29&1.2\\
19149&2275&2007ni&NTT&1.0&    1800&1.30&1.1\\
19155&2252&2007mn&NTT&1.0&     900&1.48&0.9\\
19155&2607&2007mn&NOT&1.3&    1500&1.16&0.9\\
19155&2720&2007mn&NTT&1.0&    1200&1.65&1.7\\
19221&2274&&NTT&1.0&    1800&1.19&1.0\\
19222&2299&2007nl&NTT&1.0&    1800&1.15&0.7\\
19230&2282&2007mo&NTT&1.0&    1800&1.24&1.2\\
19282&2280&2007mk&NTT&1.0&    1800&1.19&1.1\\
19323&2296&2007nc&NTT&1.0&    1800&1.20&0.8\\
19341&2298&2007nf&NTT&1.0&    1800&1.39&1.1\\
19353&2281&2007nj&NTT&1.0&    1800&1.31&1.2\\
19381&2283&2007nk&NTT&1.0&    1800&1.36&1.2\\
19899&2550&2007pu&NOT&1.3&    1800&1.14&0.9\\
19913&2585&2007qf&NTT&1.0&    1800&1.45&0.9\\
19953&2602&2007pf&NOT&1.3&    1800&1.13&0.9\\
19968&2549&2007ol&NOT&1.3&    1200&1.30&1.0\\
20039&2584&2007qh&NTT&1.0&    1800&1.44&1.5\\
20040&2612&2007rf&NTT&1.0&    1800&1.76&0.8\\
20052&2537&&NOT&1.3&    1800&1.13&0.8\\
20052&2538&&NTT&1.0&    1200&1.80&1.5\\
20088&2546&&NTT&1.0&    1800&1.59&1.0\\
20097&2587&2007rd&NTT&1.0&    1800&2.04&1.1\\
20142&2586&2007qg&NTT&1.0&    3600&1.60&0.9\\
20144&2541&2007ql&NTT&1.0&    1800&2.19&1.0\\
20227&2568&2007qi&NTT&1.0&    1800&1.39&1.2\\
20345&2567&2007qp&NTT&1.0&    1800&1.88&1.1\\
20364&2581&2007qo&NTT&1.0&    1800&2.33&1.0\\
20376&2582&2007re&NTT&1.0&    1800&1.68&0.8\\
20388&2611&&NTT&1.0&    1800&1.64&1.1\\
20430&2543&2007qj&NTT&1.0&    1800&2.03&0.9\\
20474&2563&2007rg&NTT&1.0&    1800&1.62&0.6\\
20474&2714&2007rg&NTT&1.5&    1800&2.73&1.5\\
20474&3003&2007rg&NTT&1.5&    1800&3.49&2.6\\
20530&2547&&NTT&1.0&    1200&1.50&1.3\\
20530&2571&&NTT&1.0&    1800&1.38&1.0\\
20575&2540&2007rh&NOT&1.3&    1800&1.15&1.2\\
20575&3005&2007rh&NOT&1.3&    1800&1.40&1.4\\
20625&2551&2007px&NOT&1.3&    1800&1.14&0.9\\
20625&2604&2007px&NOT&1.3&    1800&1.14&0.9\\
20677&2536&2007qx&NTT&1.0&    1800&1.64&1.8\\
20678&2610&&NTT&1.0&    1800&2.47&1.4\\
20687&2596&2007ri&NTT&1.0&    1800&1.89&1.3\\
20687&2597&2007ri&NOT&1.3&    3600&1.14&1.3\\
20718&2577&2007rj&NTT&1.0&    1300&3.15&1.0\\
20718&2593&2007rj&NTT&1.0&    1800&1.93&1.1\\
20764&2594&2007ro&NOT&1.3&    1800&1.15&1.3\\
20834&2598&2007rr&NTT&1.0&    1800&2.05&1.2\\
20862&2600&2007rn&NTT&1.0&    1800&1.49&1.1\\
20909&2580&&NTT&1.0&    1800&2.38&1.2\\
20978&2609&2007rl&NTT&1.0&    1800&2.16&1.0\\
21006&2566&2007qs&NTT&1.0&    1800&1.66&0.9\\
21033&2565&2007qy&NTT&1.0&    1800&1.61&0.8\\
21034&2719&2007qa&NTT&1.0&    1200&1.53&1.1\\
21034&2733&2007qa&NTT&1.5&    1800&1.90&2.0\\
21042&2564&2007qz&NTT&1.0&    1800&1.61&1.1\\
21058&2579&&NTT&1.0&    1800&1.62&0.7\\
21058&2595&&NOT&1.3&    1800&1.18&1.0\\
21062&2613&2007rp&NTT&1.0&    1800&2.00&1.3\\
21064&2532&2007qb&NOT&1.3&    1800&1.28&1.2\\
21064&2533&2007qb&NTT&1.0&    1200&1.57&0.9\\
21362&2636&2007sd&NOT&1.3&    2400&1.13&0.7\\
21362&2697&2007sd&NTT&1.0&    1800&1.76&1.1\\
21422&2599&2007rq&NTT&1.0&    1800&1.73&1.0\\
21502&2574&2007ra&NOT&1.3&    2076&1.13&0.9\\
21502&2575&2007ra&NOT&1.3&    1800&1.13&1.4\\
21502&2717&2007ra&NTT&1.5&    1200&1.83&1.9\\
21596&2588&&NOT&1.3&    1800&1.19&1.1\\
21596&2589&&NTT&1.0&    1800&1.57&0.7\\
21669&2591&2007rs&NTT&1.0&    1800&1.43&0.9\\
21669&2722&2007rs&NTT&1.0&     900&1.29&1.6\\
21766&2638&2007rc&NOT&1.3&    2400&1.14&0.9\\
21810&2724&2007se&NTT&1.0&    1200&2.82&1.8\\
21814&2702&2007sf&NTT&1.5&    1200&2.27&2.1\\
21839&2716&2007sl&NTT&1.5&    1200&2.27&2.0\\
21861&2723&2007sg&NTT&1.0&    1200&2.19&1.9\\
21898&2704&2007sj&NTT&1.0&    1200&2.07&1.3\\
22182&2690&2007sm&NTT&1.5&    1800&2.52&2.3\\
22284&2735&2007sn&NTT&1.0&    1200&2.49&2.6\\
\end{longtable}
\tablefoot{
\tablefoottext{a}{Most of the spectra in the table that lack IAU names are photometric SNe without reliable spectroscopic confirmation and unclassified objects.}
\tablefoottext{b}{The FWHM seeing at 5000 {\AA} for the airmass at which the observations were conducted.}
}
}

\longtab{2}{
\begin{longtable}{p{0.748cm}p{0.648cm}p{1.99cm}p{1.5cm}p{2.24cm}p{1.5cm}p{2.8cm}p{2.6cm}p{1.4cm}}
\caption{\label{tab:type} Spectroscopic typing and redshift determination.} \\
\hline \hline
ID & SPID & SDSS Type \tablefootmark{a} & NTT/NOT Type \tablefootmark{b} & LC Epoch \tablefootmark{c} & SNID Epoch \tablefootmark{d} & SDSS z \tablefootmark{e} & SNID z & Notes \tablefootmark{f}\\
\hline
\endfirsthead
\caption{continued.}\\
\hline\hline
ID & SPID & SDSS Type \tablefootmark{a}& NTT/NOT Type \tablefootmark{b}& LC Epoch \tablefootmark{c}& SNID Epoch \tablefootmark{d}& SDSS z \tablefootmark{e}& SNID z & Notes \tablefootmark{f}\\
\hline
\endhead
\hline
\multicolumn{8}{l}{{Continued on next page\ldots}} \\
\endfoot
\hline
\endlastfoot
12778&692&$-$&$-$&$-$&$-$&0.0992 $\pm$ 0.0001&$-$&\\
12779&693&Ia&Ia-norm&23.0 $\pm$ 0.8$^{p}$&26 $\pm$ 16&0.0800 $\pm$ 0.0001&
0.078 $\pm$ 0.003&s\\
12781&680&Ia&Ia-norm&10.9 $\pm$ 0.2&20 $\pm$ 18&0.0843 $\pm$ 0.0002&
0.084 $\pm$ 0.004&\\
12782&681&II&II&$-$&$-$&0.06787 $\pm$ 0.00005&$-$&v\\
12820&711&II&IIP&$-$&18 $\pm$ 66&0.04458 $\pm$ 0.00005&0.046 $\pm$ 0.005&\\
12842&682&II&II&$-$&$-$&0.0887 $\pm$ 0.0005&$-$&v, zg\\
12843&727&Ia&Ia-norm&10.2 $\pm$ 0.1&16 $\pm$ 25&0.1670 $\pm$ 0.0001&
0.161 $\pm$ 0.006&s\\
12844&684&Ic&$-$&$-$&$-$&0.07053 $\pm$ 0.00009&$-$&\\
12853&685&Ia&Ia-norm&10.3 $\pm$ 0.2&11 $\pm$ 52&0.1694 $\pm$ 0.0005&
0.171 $\pm$ 0.006&zg\\
12855&716&Ia&Ia-norm&-2.6 $\pm$ 0.2&0 $\pm$ 5&0.172 $\pm$ 0.005&
0.171 $\pm$ 0.005&s, zs\\
12856&695&Ia&Ia-norm&-3.2 $\pm$ 0.2&-1 $\pm$ 4&0.1717 $\pm$ 0.0001&
0.171 $\pm$ 0.005&s\\
12860&688&Ia&Ia-norm&-1.9 $\pm$ 0.9&-1 $\pm$ 5&0.1217 $\pm$ 0.0005&
0.124 $\pm$ 0.005&s, zg\\
12874&689&$-$&$-$&$-$&$-$&0.2449 $\pm$ 0.0002&$-$&\\
12898&712&Ia&Ia-norm&-6.6 $\pm$ 0.1&-1 $\pm$ 5&0.0835 $\pm$ 0.0005&
0.078 $\pm$ 0.004&s, zg\\
12907&714&Ia&Ia-norm&-0.1 $\pm$ 0.2&0 $\pm$ 5&0.1318 $\pm$ 0.0002&
0.124 $\pm$ 0.005&s\\
12927&690&Ia&Ia-norm&2.6 $\pm$ 0.8$^{p}$&-1 $\pm$ 4&0.175 $\pm$ 0.005&
0.172 $\pm$ 0.005&s, zs\\
12928&686&Ia&Ia-norm&17 $\pm$ 1$^{p}$&20 $\pm$ 24&0.1397 $\pm$ 0.0005&
0.141 $\pm$ 0.004&s, zg\\
12930&687&Ia&Ia-norm&10.1 $\pm$ 0.2&12 $\pm$ 12&0.1475 $\pm$ 0.0002&
0.140 $\pm$ 0.005&\\
12947&691&photo-Ia&$-$&1.4 $\pm$ 0.2&$-$&0.1592 $\pm$ 0.0005&$-$&zg\\
12950&700&Ia&Ia-norm&-4.43 $\pm$ 0.04&-1 $\pm$ 4&0.08268 $\pm$ 0.00004&
0.085 $\pm$ 0.004&s\\
12950&1055&Ia&Ia-norm&22.41 $\pm$ 0.04&32 $\pm$ 9&0.08268 $\pm$ 0.00004&
0.085 $\pm$ 0.002&s\\
12978&701&photo-Ia&$-$&$-$&$-$&$-$&$-$&v\\
13005&702&Ia&Ia-norm&24 $\pm$ 1$^{p}$&31 $\pm$ 11&0.1273 $\pm$ 0.0005&
0.127 $\pm$ 0.003&s, zg\\
13025&761&Ia&Ia-norm&3.4 $\pm$ 0.3&2 $\pm$ 5&0.2239 $\pm$ 0.0005&
0.228 $\pm$ 0.006&zg\\
13044&724&Ia&Ia-norm&-8.20 $\pm$ 0.06&-1 $\pm$ 4&0.1257 $\pm$ 0.0005&
0.126 $\pm$ 0.004&zg\\
13044&1062&Ia&Ia-norm&20.22 $\pm$ 0.06&24 $\pm$ 18&0.1257 $\pm$ 0.0005&
0.126 $\pm$ 0.004&s, zg\\
13045&734&Ia&Ia&0.5 $\pm$ 0.3&1 $\pm$ 5&0.1808 $\pm$ 0.0005&0.183 $\pm$ 0.006&
s, zg\\
13046&726&$-$&Gal&$-$&$-$&0.1259 $\pm$ 0.0005&0.1245 $\pm$ 0.0009&zg, f\\
13070&736&Ia&Ia&6.9 $\pm$ 0.2&8 $\pm$ 2&0.19855 $\pm$ 0.00009&0.195 $\pm$ 0.007&
\\
13072&723&Ia&Ia-norm&0 $\pm$ 2&-1 $\pm$ 5&0.2306 $\pm$ 0.0008&0.234 $\pm$ 0.005&
s\\
13135&739&Ia&Ia-norm&-7.74 $\pm$ 0.07&-1 $\pm$ 5&0.1047 $\pm$ 0.0001&
0.098 $\pm$ 0.005&\\
13135&998&Ia&Ia-norm&17.64 $\pm$ 0.07&32 $\pm$ 12&0.1047 $\pm$ 0.0001&
0.101 $\pm$ 0.002&\\
13174&766&Ia?&$-$&-3.8 $\pm$ 0.1&$-$&0.2361 $\pm$ 0.0005&$-$&v, zg\\
13195&764&Ib&Ib&$-$&6 $\pm$ 16&0.0207 $\pm$ 0.0001&0.013 $\pm$ 0.006&\\
13195&983&Ib&Ib-norm&$-$&14 $\pm$ 21&0.0207 $\pm$ 0.0001&0.022 $\pm$ 0.007&\\
13195&1458&Ib&Ib&$-$&$-$&0.0207 $\pm$ 0.0001&$-$&v\\
13355&1003&II&IIP&$-$&35 $\pm$ 105&0.05969 $\pm$ 0.00008&0.060 $\pm$ 0.004&\\
13376&1002&II&IIP&$-$&33 $\pm$ 76&0.0698 $\pm$ 0.0001&0.069 $\pm$ 0.005&\\
13376&1106&II&IIP&$-$&34 $\pm$ 83&0.0698 $\pm$ 0.0001&0.069 $\pm$ 0.004&\\
13577&1000&AGN&Gal&$-$&$-$&0.2309 $\pm$ 0.0005&0.2309 $\pm$ 0.0006&zg\\
13796&1058&Ia&Ia-norm&13 $\pm$ 3&15 $\pm$ 42&0.1482 $\pm$ 0.0005&
0.149 $\pm$ 0.006&zg\\
13894&1039&Ia&Ia-norm&9.2 $\pm$ 0.1&8 $\pm$ 5&0.1249 $\pm$ 0.0005&
0.127 $\pm$ 0.006&s, zg\\
14157&1040&Ia&Ia&9.4 $\pm$ 0.2&9 $\pm$ 5&0.2115 $\pm$ 0.0005&0.210 $\pm$ 0.005&
s, zg\\
14279&1459&Ia&NotSN&41.30 $\pm$ 0.03&$-$&0.0454 $\pm$ 0.0002&0.0461 $\pm$ 0.0009
&s\\
14318&1594&Ia&Ia-norm&-4.3 $\pm$ 0.3$^{p}$&-3 $\pm$ 5&0.0579 $\pm$ 0.0002&
0.054 $\pm$ 0.004&s\\
14318&1653&Ia&Ia&11.7 $\pm$ 0.3$^{p}$&12 $\pm$ 12&0.0579 $\pm$ 0.0002&
0.056 $\pm$ 0.005&s\\
14318&1713&Ia&Ia&13.6 $\pm$ 0.3$^{p}$&16 $\pm$ 10&0.0579 $\pm$ 0.0002&
0.057 $\pm$ 0.004&s\\
14437&1061&Ia&Ia&14.0 $\pm$ 0.1&17 $\pm$ 41&0.1491 $\pm$ 0.0005&
0.148 $\pm$ 0.005&s, zg\\
14450&991&II&IIP&$-$&15 $\pm$ 9&0.1203 $\pm$ 0.0001&0.117 $\pm$ 0.003&\\
14451&989&Ia&Ia-norm&9.5 $\pm$ 0.1&8 $\pm$ 2&0.1784 $\pm$ 0.0005&
0.1813 $\pm$ 0.0010&s, zg\\
14492&1001&Ib&Ib/c&$-$&$-$&0.0767 $\pm$ 0.0001&$-$&v\\
14598&987&$-$&$-$&$-$&$-$&$-$&$-$&\\
14599&988&II&IIP&$-$&5 $\pm$ 25&0.0555 $\pm$ 0.0005&0.048 $\pm$ 0.005&zg\\
14782&990&Ia&Ia-norm&1.7 $\pm$ 0.1&-1 $\pm$ 5&0.1604 $\pm$ 0.0005&
0.165 $\pm$ 0.005&s, zg\\
14846&1014&Ia&Ia-norm&-1.7 $\pm$ 0.2&-1 $\pm$ 4&0.2247 $\pm$ 0.0005&
0.223 $\pm$ 0.005&s, zg\\
14871&1008&Ia&Ia-norm&-4.2 $\pm$ 0.1&-1 $\pm$ 5&0.1276 $\pm$ 0.0005&
0.125 $\pm$ 0.005&s, zg\\
14979&1009&Ia&Ia-norm&-2.1 $\pm$ 0.1&-2 $\pm$ 4&0.1771 $\pm$ 0.0005&
0.178 $\pm$ 0.006&zg\\
14984&1027&Ia&Ia-norm&-1.2 $\pm$ 0.2&0 $\pm$ 5&0.1967 $\pm$ 0.0005&
0.191 $\pm$ 0.005&s, zg\\
15031&985&II&IIP&$-$&4 $\pm$ 22&0.03073 $\pm$ 0.00009&0.028 $\pm$ 0.004&\\
15129&1015&Ia&Ia-norm&1.8 $\pm$ 0.1&0 $\pm$ 4&0.1985 $\pm$ 0.0002&
0.201 $\pm$ 0.005&s\\
15132&1012&Ia&Ia-norm&-2.4 $\pm$ 0.2&0 $\pm$ 4&0.144 $\pm$ 0.005&
0.151 $\pm$ 0.005&zs\\
15136&1022&Ia?&Ia?&-0.3 $\pm$ 0.1&$-$&0.14869 $\pm$ 0.00005&$-$&v\\
15153&1046&photo-II&$-$&$-$&$-$&$-$&$-$&v\\
15161&1010&Ia&Ia-norm&-1.0 $\pm$ 0.4&-4 $\pm$ 5&0.2496 $\pm$ 0.0002&
0.253 $\pm$ 0.005&\\
15171&1045&Ia&Ia-norm&-5.7 $\pm$ 0.1&-1 $\pm$ 4&0.134 $\pm$ 0.005&
0.139 $\pm$ 0.005&zs\\
15203&1026&Ia&Ia-norm&-2.4 $\pm$ 0.2&-1 $\pm$ 5&0.2043 $\pm$ 0.0005&
0.199 $\pm$ 0.006&s, zg\\
15207&1038&$-$&$-$&$-$&$-$&0.2582 $\pm$ 0.0005&$-$&zg\\
15210&1005&photo-non-Ia&$-$&$-$&$-$&0.1262 $\pm$ 0.0005&$-$&v, zg\\
15210&1052&photo-non-Ia&$-$&$-$&$-$&0.1262 $\pm$ 0.0005&$-$&zg, f\\
15222&1004&Ia&Ia-norm&-5.8 $\pm$ 0.2&3 $\pm$ 7&0.1994 $\pm$ 0.0001&
0.1975 $\pm$ 0.0010&f\\
15234&1043&Ia&Ia-norm&-7.8 $\pm$ 0.2&-2 $\pm$ 5&0.13634 $\pm$ 0.00009&
0.134 $\pm$ 0.002&s\\
15259&1051&Ia&Ia-norm&-1.9 $\pm$ 0.2&0 $\pm$ 5&0.2100 $\pm$ 0.0001&
0.213 $\pm$ 0.007&\\
15287&1057&Ia&Ia&-3.4 $\pm$ 0.3&0 $\pm$ 5&0.254 $\pm$ 0.005&0.255 $\pm$ 0.009&zs
\\
15320&1098&II&$-$&$-$&$-$&0.062 $\pm$ 0.005&$-$&v, zs\\
15339&1107&II&$-$&$-$&$-$&0.1200 $\pm$ 0.0001&$-$&\\
15354&1110&Ia&Gal&1.3 $\pm$ 0.4&$-$&0.2221 $\pm$ 0.0005&0.2209 $\pm$ 0.0006&
zg, f\\
15475&1464&Ic&Ib/c&$-$&$-$&0.0162 $\pm$ 0.0002&$-$&v\\
15557&1532&II&IIP&$-$&2 $\pm$ 3&0.2860 $\pm$ 0.0005&0.301 $\pm$ 0.002&zg, f\\
16021&1355&Ia&Ia-norm&11.29 $\pm$ 0.09&9 $\pm$ 4&0.124 $\pm$ 0.005&
0.130 $\pm$ 0.005&s, zs\\
16069&1358&Ia&Ia&4.3 $\pm$ 0.2&$-$&0.1288 $\pm$ 0.0001&$-$&v\\
16069&1467&Ia&Ia-norm&11.5 $\pm$ 0.2&12 $\pm$ 26&0.1288 $\pm$ 0.0001&
0.124 $\pm$ 0.004&s\\
16069&1651&Ia&Ia&28.3 $\pm$ 0.2&$-$&0.1288 $\pm$ 0.0001&$-$&v\\
16087&1455&II&IIP&$-$&14 $\pm$ 23&0.0554 $\pm$ 0.0001&0.052 $\pm$ 0.005&\\
16163&1678&photo-Ia&Gal&30.6 $\pm$ 0.3&$-$&0.1549 $\pm$ 0.0005&
0.1557 $\pm$ 0.0006&zg\\
16165&1326&Ia&Ia-norm&2.6 $\pm$ 0.1&0 $\pm$ 5&0.157 $\pm$ 0.005&
0.161 $\pm$ 0.005&zs\\
16179&1323&Ic&$-$&$-$&$-$&0.1370 $\pm$ 0.0005&$-$&zg\\
16179&1469&Ic&Ic&$-$&$-$&0.1370 $\pm$ 0.0005&$-$&v, zg\\
16179&1570&Ic&Ic&$-$&-1 $\pm$ 3&0.1370 $\pm$ 0.0005&0.13 $\pm$ 0.01&zg\\
16192&1322&II&$-$&$-$&$-$&0.0787 $\pm$ 0.0001&$-$&\\
16192&1496&II&IIP&$-$&17 $\pm$ 32&0.0787 $\pm$ 0.0001&0.074 $\pm$ 0.006&\\
16204&1500&photo-non-Ia&Gal&$-$&$-$&0.0546 $\pm$ 0.0002&0.0537 $\pm$ 0.0006&f\\
16206&1501&Ia&Ia-norm&6.1 $\pm$ 0.2&11 $\pm$ 9&0.1600 $\pm$ 0.0005&
0.165 $\pm$ 0.007&s, zg\\
16215&1456&Ia&Ia-norm&4.29 $\pm$ 0.09&0 $\pm$ 6&0.0466 $\pm$ 0.0002&
0.045 $\pm$ 0.005&s\\
16215&1630&Ia&Ia-norm&24.28 $\pm$ 0.09&31 $\pm$ 8&0.0466 $\pm$ 0.0002&
0.046 $\pm$ 0.002&s\\
16241&1470&photo-non-Ia&$-$&$-$&$-$&0.0979 $\pm$ 0.0005&$-$&zg\\
16280&1471&Ia&Ia-norm&5.61 $\pm$ 0.07&8 $\pm$ 11&0.0381 $\pm$ 0.0002&
0.036 $\pm$ 0.009&s\\
16280&1564&Ia&Ia&23.89 $\pm$ 0.07&16 $\pm$ 8&0.0381 $\pm$ 0.0002&
0.037 $\pm$ 0.008&s\\
16287&1449&Ia&Ia-norm&2.4 $\pm$ 0.1&0 $\pm$ 5&0.1064 $\pm$ 0.0005&
0.106 $\pm$ 0.005&zg\\
16287&1569&Ia&Ia-norm&3.3 $\pm$ 0.1&0 $\pm$ 5&0.1064 $\pm$ 0.0005&
0.109 $\pm$ 0.005&zg\\
16287&1650&Ia&Ia-norm&19.5 $\pm$ 0.1&31 $\pm$ 8&0.1064 $\pm$ 0.0005&
0.105 $\pm$ 0.003&s, zg\\
16302&1473&photo-Ia&$-$&$-$&$-$&$-$&$-$&v\\
16314&1335&Ia&Ia&-10.5 $\pm$ 0.2$^{p}$&-10 $\pm$ 6&0.0626 $\pm$ 0.0001&
0.072 $\pm$ 0.006&f\\
16314&1475&Ia&Ia&-4.9 $\pm$ 0.2$^{p}$&-2 $\pm$ 5&0.0626 $\pm$ 0.0001&
0.063 $\pm$ 0.006&\\
16333&1367&Ia&Ia-91T&-7.4 $\pm$ 0.4&-7 $\pm$ 3&0.0719 $\pm$ 0.0001&
0.067 $\pm$ 0.002&s\\
16352&1478&Ia&Ia-norm&4.1 $\pm$ 0.4&1 $\pm$ 4&0.248 $\pm$ 0.005&
0.255 $\pm$ 0.007&zs\\
16391&1452&II&$-$&$-$&$-$&0.12 $\pm$ 0.01&$-$&zt\\
16391&1565&II&II&$-$&0 $\pm$ 38&0.12 $\pm$ 0.01&0.109 $\pm$ 0.007&zt\\
16392&1365&Ia&Ia-norm&-8.4 $\pm$ 0.5$^{p}$&-1 $\pm$ 5&0.0592 $\pm$ 0.0002&
0.057 $\pm$ 0.005&s\\
16392&1448&Ia&Ia-norm&-1.8 $\pm$ 0.5$^{p}$&-1 $\pm$ 5&0.0592 $\pm$ 0.0002&
0.061 $\pm$ 0.005&\\
16392&1566&Ia&NotSN&16.2 $\pm$ 0.5$^{p}$&$-$&0.0592 $\pm$ 0.0002&
0.0583 $\pm$ 0.0009&s\\
16392&1682&Ia&Ia&18.0 $\pm$ 0.5$^{p}$&24 $\pm$ 12&0.0592 $\pm$ 0.0002&
0.053 $\pm$ 0.007&s\\
16402&1505&Ia&$-$&3.5 $\pm$ 0.3&$-$&0.2645 $\pm$ 0.0005&$-$&s, zg\\
16473&1520&Ia&Ia-norm&1.3 $\pm$ 0.9&0 $\pm$ 5&0.211 $\pm$ 0.005&
0.215 $\pm$ 0.006&s, zs\\
16541&1485&Ia&Ia&-5 $\pm$ 1$^{p}$&1 $\pm$ 3&0.128 $\pm$ 0.005&0.131 $\pm$ 0.004&
zs\\
16578&1516&Ia&Ia-norm&2.9 $\pm$ 0.9$^{p}$&0 $\pm$ 5&0.1747 $\pm$ 0.0005&
0.178 $\pm$ 0.006&s, zg\\
16619&1519&Ia&Ia-norm&-3.6 $\pm$ 0.5$^{p}$&0 $\pm$ 4&0.101 $\pm$ 0.005&
0.106 $\pm$ 0.004&zs\\
16619&1528&Ia&Ia-norm&-6.6 $\pm$ 0.5$^{p}$&0 $\pm$ 4&0.101 $\pm$ 0.005&
0.104 $\pm$ 0.005&zs\\
16637&1514&Ia&Ia-norm&-1 $\pm$ 1&-6 $\pm$ 9&0.2430 $\pm$ 0.0005&0.24 $\pm$ 0.01&
zg, f\\
16641&1518&Ia&Ia&-3.2 $\pm$ 0.5$^{p}$&$-$&0.1265 $\pm$ 0.0005&$-$&v, zg\\
16641&1530&Ia&Ia-norm&-6.1 $\pm$ 0.5$^{p}$&-1 $\pm$ 5&0.1265 $\pm$ 0.0005&
0.118 $\pm$ 0.005&s, zg\\
16641&1649&Ia&Ia-norm&11.9 $\pm$ 0.5$^{p}$&14 $\pm$ 39&0.1265 $\pm$ 0.0005&
0.123 $\pm$ 0.005&s, zg\\
16668&1561&II&IIn&$-$&318 $\pm$ 100&0.1518 $\pm$ 0.0005&0.169 $\pm$ 0.005&zg\\
16692&1489&Ia&Ia-norm&-6.5 $\pm$ 0.3$^{p}$&-1 $\pm$ 4&0.0341 $\pm$ 0.0002&
0.033 $\pm$ 0.003&\\
16737&1599&Ia&Ia-norm&-5 $\pm$ 1$^{p}$&-2 $\pm$ 5&0.200 $\pm$ 0.005&
0.191 $\pm$ 0.007&zs\\
16741&1523&$-$&$-$&$-$&$-$&$-$&$-$&\\
16748&1574&Ia&Gal&0 $\pm$ 2$^{p}$&$-$&0.2320 $\pm$ 0.0005&0.232 $\pm$ 0.004&
zg, f\\
16774&1606&Ia&$-$&-9 $\pm$ 2$^{p}$&$-$&0.2146 $\pm$ 0.0005&$-$&zg, f\\
16778&1542&$-$&$-$&$-$&$-$&0.0861 $\pm$ 0.0005&$-$&zg\\
16778&1568&$-$&$-$&$-$&$-$&0.0861 $\pm$ 0.0005&$-$&zg\\
16793&1603&Ia&Ia-norm&-2.5 $\pm$ 0.7$^{p}$&-1 $\pm$ 6&0.222 $\pm$ 0.005&
0.221 $\pm$ 0.009&zs\\
16838&1522&II&IIn (sn98S)&$-$&-12.8 $\pm$ 0.4&0.15 $\pm$ 0.01&
0.14945 $\pm$ 0.00007&zt, f\\
16857&1538&$-$&$-$&$-$&$-$&0.0753 $\pm$ 0.0002&$-$&\\
16867&1541&$-$&$-$&$-$&$-$&0.1195 $\pm$ 0.0005&$-$&zg\\
16872&1539&Ia&Ia-norm&-4.6 $\pm$ 1.0$^{p}$&-1 $\pm$ 4&0.1266 $\pm$ 0.0005&
0.119 $\pm$ 0.004&zg\\
16956&1562&Ia&Ia-norm&-0.2 $\pm$ 0.7$^{p}$&1 $\pm$ 6&0.1087 $\pm$ 0.0005&
0.107 $\pm$ 0.006&zg\\
16979&1597&photo-non-Ia&$-$&$-$&$-$&0.4874 $\pm$ 0.0005&$-$&zg\\
16988&1595&Ib&$-$&$-$&$-$&0.05836 $\pm$ 0.00008&$-$&\\
16988&1652&Ib&Ib-norm&$-$&12 $\pm$ 13&0.05836 $\pm$ 0.00008&0.058 $\pm$ 0.007&\\
17117&1679&Ia&Ia-norm&7.4 $\pm$ 0.5$^{p}$&-1 $\pm$ 5&0.14017 $\pm$ 0.00004&
0.143 $\pm$ 0.005&\\
17135&1648&Ia&Ia-norm&$-$&-1 $\pm$ 5&0.03092 $\pm$ 0.00008&0.032 $\pm$ 0.004&\\
17167&2250&II&IIP&$-$&68 $\pm$ 44&0.0849 $\pm$ 0.0005&0.076 $\pm$ 0.003&zg\\
17170&1879&$-$&$-$&$-$&$-$&0.2205 $\pm$ 0.0005&$-$&zg\\
17176&1812&Ia&Ia&16 $\pm$ 2$^{p}$&26 $\pm$ 9&0.0935 $\pm$ 0.0001&
0.105 $\pm$ 0.007&s\\
17200&1796&II&IIP&$-$&48 $\pm$ 50&0.08469 $\pm$ 0.00009&0.084 $\pm$ 0.003&\\
17206&1788&photo-Ia&$-$&9.0 $\pm$ 0.4&$-$&0.1564 $\pm$ 0.0001&$-$&\\
17218&1794&Ia&Ia&10.5 $\pm$ 0.5&$-$&0.1783 $\pm$ 0.0005&$-$&v, zg\\
17220&1791&Ia&Ia-norm&4.2 $\pm$ 0.2&2 $\pm$ 6&0.1783 $\pm$ 0.0005&
0.179 $\pm$ 0.007&s, zg\\
17223&1793&Ia&Ia-norm&13 $\pm$ 1$^{p}$&11 $\pm$ 51&0.235 $\pm$ 0.005&
0.236 $\pm$ 0.007&zs\\
17237&1830&photo-non-Ia&NotSN&$-$&$-$&0.2516 $\pm$ 0.0005&0.253 $\pm$ 0.001&zg\\
17245&2234&photo-non-Ia&Gal&$-$&$-$&0.2784 $\pm$ 0.0005&0.2784 $\pm$ 0.0004&zg\\
17247&1799&photo-Ia&Gal&23 $\pm$ 2$^{p}$&$-$&0.2021 $\pm$ 0.0005&
0.2018 $\pm$ 0.0007&zg\\
17253&1898&Ia&Ia-norm&18.1 $\pm$ 0.9$^{p}$&21 $\pm$ 27&0.1560 $\pm$ 0.0005&
0.158 $\pm$ 0.005&s, zg\\
17254&1813&Ia?&Ia?&6 $\pm$ 4$^{p}$&$-$&0.2691 $\pm$ 0.0005&$-$&v, zg\\
17332&1899&Ia&Ia-norm&3.3 $\pm$ 0.2&2 $\pm$ 6&0.1828 $\pm$ 0.0001&
0.183 $\pm$ 0.007&\\
17351&1769&$-$&$-$&$-$&$-$&0.1832 $\pm$ 0.0007&$-$&zg\\
17366&1782&Ia&Ia-norm&8.9 $\pm$ 0.2&6 $\pm$ 53&0.1393 $\pm$ 0.0002&
0.143 $\pm$ 0.008&\\
17389&1811&Ia&Ia&7.0 $\pm$ 0.3&6 $\pm$ 5&0.1706 $\pm$ 0.0005&0.171 $\pm$ 0.007&
s, zg\\
17391&1872&Ia&Ia-norm&16 $\pm$ 2$^{p}$&20 $\pm$ 26&0.1849 $\pm$ 0.0005&
0.193 $\pm$ 0.005&zg\\
17422&1785&II&II&$-$&$-$&0.1493 $\pm$ 0.0005&$-$&v, zg\\
17435&1902&Ia&Ia-norm&2.7 $\pm$ 0.2&1 $\pm$ 4&0.2180 $\pm$ 0.0005&
0.224 $\pm$ 0.006&zg\\
17436&1790&$-$&$-$&$-$&$-$&0.1449 $\pm$ 0.0006&$-$&v, zg\\
17464&1853&Ia&Ia&3 $\pm$ 1$^{p}$&7 $\pm$ 5&0.2549 $\pm$ 0.0005&0.250 $\pm$ 0.003
&zg\\
17486&1854&photo-non-Ia&$-$&$-$&$-$&0.4476 $\pm$ 0.0002&$-$&f\\
17497&1837&Ia&Ia-norm&-2.38 $\pm$ 0.09&-1 $\pm$ 5&0.1448 $\pm$ 0.0001&
0.146 $\pm$ 0.005&s\\
17500&2249&Ia&Ia-norm&67.9 $\pm$ 0.1$^{p}$&66 $\pm$ 35&0.0441 $\pm$ 0.0002&
0.043 $\pm$ 0.005&\\
17535&1838&photo-II&$-$&$-$&$-$&$-$&$-$&\\
17548&1825&Ic&Gal&$-$&$-$&0.0393 $\pm$ 0.0005&0.0392 $\pm$ 0.0003&zg\\
17548&2231&Ic&Ic&$-$&$-$&0.0393 $\pm$ 0.0005&$-$&v, zg\\
17548&2293&Ic&Ic&$-$&$-$&0.0393 $\pm$ 0.0005&$-$&v, zg\\
17552&1789&Ia&Ia-norm&3.5 $\pm$ 0.5&0 $\pm$ 5&0.2542 $\pm$ 0.0005&
0.255 $\pm$ 0.007&s, zg\\
17568&1810&Ia&Ia-norm&1 $\pm$ 4&-1 $\pm$ 4&0.1445 $\pm$ 0.0005&0.141 $\pm$ 0.004
&s, zg\\
17605&1809&Ia&Ia&8 $\pm$ 2&$-$&0.1465 $\pm$ 0.0005&$-$&v, zg\\
17627&1841&II&IIP&$-$&5 $\pm$ 60&0.06966 $\pm$ 0.00008&0.070 $\pm$ 0.003&\\
17629&1851&Ia&Ia-norm&-7.13 $\pm$ 0.07&1 $\pm$ 6&0.13690 $\pm$ 0.00007&
0.136 $\pm$ 0.006&s\\
17647&1875&photo-Ia&$-$&$-$&$-$&$-$&$-$&v\\
17703&1881&photo-II&$-$&$-$&$-$&$-$&$-$&\\
17745&2161&Ia&Ia&15.11 $\pm$ 0.08&16 $\pm$ 32&0.0636 $\pm$ 0.0005&
0.063 $\pm$ 0.005&s, zg\\
17746&1873&$-$&$-$&$-$&$-$&0.157 $\pm$ 0.005&$-$&zs\\
17784&1842&Ia&Ia-norm&-5.55 $\pm$ 0.03&-1 $\pm$ 5&0.03710 $\pm$ 0.00007&
0.029 $\pm$ 0.005&\\
17790&1887&Ia&Ia-norm&1.0 $\pm$ 0.4&0 $\pm$ 5&0.178 $\pm$ 0.005&
0.177 $\pm$ 0.006&zs\\
17794&1906&photo-Ibc&$-$&$-$&$-$&$-$&$-$&\\
17811&1816&Ia&Ia&4.6 $\pm$ 0.8&1 $\pm$ 5&0.2132 $\pm$ 0.0005&0.205 $\pm$ 0.006&
zg\\
17814&1901&photo-non-Ia&Gal&$-$&$-$&0.1069 $\pm$ 0.0005&0.1071 $\pm$ 0.0005&zg\\
17825&1819&Ia&Ia-norm&-4.9 $\pm$ 0.1&-1 $\pm$ 5&0.161 $\pm$ 0.005&
0.162 $\pm$ 0.005&zs\\
17854&2230&$-$&$-$&$-$&$-$&$-$&$-$&v\\
17875&1817&Ia&Ia-norm&0.3 $\pm$ 0.3&0 $\pm$ 5&0.2323 $\pm$ 0.0005&
0.223 $\pm$ 0.008&s, zg\\
17880&1843&Ia&Ia-norm&-1.87 $\pm$ 0.06&0 $\pm$ 6&0.07265 $\pm$ 0.00006&
0.061 $\pm$ 0.005&\\
17880&1957&Ia&Ia-norm&1.16 $\pm$ 0.06&-1 $\pm$ 6&0.07265 $\pm$ 0.00006&
0.065 $\pm$ 0.005&s\\
17880&2194&Ia&Ia&24.44 $\pm$ 0.06&30 $\pm$ 7&0.07265 $\pm$ 0.00006&
0.072 $\pm$ 0.005&s\\
17886&1844&Ia&Ia-norm&-4.48 $\pm$ 0.04&-1 $\pm$ 5&0.0408 $\pm$ 0.0002&
0.040 $\pm$ 0.005&\\
17924&1826&photo-non-Ia&Gal&$-$&$-$&0.1456 $\pm$ 0.0005&0.1444 $\pm$ 0.0008&
zg, f\\
17973&1926&photo-non-Ia&NotSN&$-$&$-$&0.1456 $\pm$ 0.0005&0.149 $\pm$ 0.001&zg\\
17973&1942&photo-non-Ia&Gal&$-$&$-$&0.1456 $\pm$ 0.0005&0.148 $\pm$ 0.001&zg, f
\\
18109&1940&II&$-$&$-$&$-$&0.0680 $\pm$ 0.0002&$-$&\\
18325&2277&Ia&Ia-norm&8.6 $\pm$ 0.3&10 $\pm$ 51&0.255 $\pm$ 0.005&
0.259 $\pm$ 0.008&zs\\
18457&2285&II&$-$&$-$&$-$&0.08097 $\pm$ 0.00008&$-$&\\
18466&2270&Ia&Ia-norm&4.5 $\pm$ 0.3&0 $\pm$ 5&0.213 $\pm$ 0.005&
0.218 $\pm$ 0.006&zs\\
18590&2248&II&$-$&$-$&$-$&0.0572 $\pm$ 0.0002&$-$&\\
18596&2227&II&IIP&$-$&7 $\pm$ 21&0.027 $\pm$ 0.005&0.023 $\pm$ 0.004&zs\\
18647&2271&photo-non-Ia&$-$&$-$&$-$&0.2128 $\pm$ 0.0002&$-$&\\
18697&2171&Ia&Ia&4.27 $\pm$ 0.08&8 $\pm$ 3&0.10725 $\pm$ 0.00005&
0.106 $\pm$ 0.005&s\\
18768&2135&Ia&Ia-norm&6.7 $\pm$ 0.3&6 $\pm$ 5&0.198 $\pm$ 0.005&
0.199 $\pm$ 0.006&zs\\
18787&2150&Ia&Ia-norm&0.2 $\pm$ 0.4&-1 $\pm$ 4&0.2073 $\pm$ 0.0005&
0.196 $\pm$ 0.006&zg\\
18804&2148&Ia&Ia-norm&-4.99 $\pm$ 0.09&-1 $\pm$ 6&0.2052 $\pm$ 0.0005&
0.194 $\pm$ 0.007&zg\\
18903&2247&Ia?&Ia?&-3.9 $\pm$ 0.3&$-$&0.1564 $\pm$ 0.0002&$-$&v\\
18965&2279&Ia&Ia-norm&-4.3 $\pm$ 0.2&-1 $\pm$ 4&0.2066 $\pm$ 0.0005&
0.207 $\pm$ 0.005&s, zg\\
19003&2235&Ia&Ia-norm&-6.4 $\pm$ 0.1&-5 $\pm$ 4&0.0612 $\pm$ 0.0002&
0.056 $\pm$ 0.005&s\\
19003&2290&Ia&Ia-norm&-4.6 $\pm$ 0.1&-6 $\pm$ 5&0.0612 $\pm$ 0.0002&
0.060 $\pm$ 0.004&s\\
19008&2284&Ia&Ia-norm&-2.1 $\pm$ 0.3&-6 $\pm$ 8&0.2322 $\pm$ 0.0005&
0.230 $\pm$ 0.009&zg\\
19023&2236&Ia&Ia-norm&-1.7 $\pm$ 0.2&-2.9 $\pm$ 0.5&0.243 $\pm$ 0.005&
0.264 $\pm$ 0.007&zs, f\\
19051&2297&$-$&$-$&$-$&$-$&0.2773 $\pm$ 0.0005&$-$&zg, f\\
19101&2268&Ia&Ia-norm&-6.0 $\pm$ 0.1&0 $\pm$ 5&0.187 $\pm$ 0.005&
0.189 $\pm$ 0.006&zs\\
19149&2275&Ia&Ia-91T&-7.1 $\pm$ 0.1&-6 $\pm$ 2&0.196 $\pm$ 0.005&
0.204 $\pm$ 0.001&zs\\
19155&2252&Ia&Ia-norm&-11.66 $\pm$ 0.04&-5 $\pm$ 4&0.0769 $\pm$ 0.0002&
0.070 $\pm$ 0.001&\\
19155&2607&Ia&Ia&18.69 $\pm$ 0.04&16 $\pm$ 29&0.0769 $\pm$ 0.0002&
0.077 $\pm$ 0.004&s\\
19155&2720&Ia&Ia-norm&38.37 $\pm$ 0.04&34 $\pm$ 12&0.0769 $\pm$ 0.0002&
0.079 $\pm$ 0.002&s\\
19221&2274&$-$&$-$&$-$&$-$&0.0433 $\pm$ 0.0005&$-$&v, zg\\
19222&2299&II&IIP&$-$&0 $\pm$ 2&0.1683 $\pm$ 0.0005&0.162 $\pm$ 0.004&zg\\
19230&2282&Ia&Ia-norm&-1.9 $\pm$ 0.2&1 $\pm$ 5&0.2215 $\pm$ 0.0005&
0.223 $\pm$ 0.006&s, zg\\
19282&2280&Ia&Ia-norm&-8.16 $\pm$ 0.10&-2 $\pm$ 5&0.1864 $\pm$ 0.0002&
0.177 $\pm$ 0.006&\\
19323&2296&Ib&Ib-norm&$-$&11 $\pm$ 14&0.08679 $\pm$ 0.00009&0.073 $\pm$ 0.007&\\
19341&2298&Ia&Ia-norm&-2.4 $\pm$ 0.3&-1 $\pm$ 5&0.228 $\pm$ 0.005&
0.229 $\pm$ 0.005&s, zs\\
19353&2281&Ia&Ia-norm&-7 $\pm$ 2&-2 $\pm$ 5&0.1540 $\pm$ 0.0001&
0.149 $\pm$ 0.005&s\\
19381&2283&Ia&Ia-norm&-3.49 $\pm$ 0.07&0 $\pm$ 4&0.210 $\pm$ 0.005&
0.212 $\pm$ 0.005&zs\\
19899&2550&Ia&Ia-norm&1.22 $\pm$ 0.09&0 $\pm$ 4&0.089 $\pm$ 0.005&
0.092 $\pm$ 0.003&zs\\
19913&2585&Ia&Ia-norm&9.6 $\pm$ 0.3&9 $\pm$ 15&0.2038 $\pm$ 0.0005&
0.208 $\pm$ 0.006&zg\\
19953&2602&Ia&Ia-norm&4.2 $\pm$ 0.1&0 $\pm$ 5&0.120 $\pm$ 0.005&
0.124 $\pm$ 0.005&zs\\
19968&2549&Ia&Ia-norm&5.24 $\pm$ 0.05&1 $\pm$ 6&0.0560 $\pm$ 0.0001&
0.059 $\pm$ 0.005&\\
20039&2584&Ia&Ia-norm&7.6 $\pm$ 0.3&6 $\pm$ 5&0.2477 $\pm$ 0.0005&
0.250 $\pm$ 0.007&s, zg\\
20040&2612&Ia&Ia&6.4 $\pm$ 0.5&-2 $\pm$ 5&0.2880 $\pm$ 0.0005&0.288 $\pm$ 0.006&
s, zg\\
20052&2537&photo-non-Ia&NotSN&$-$&$-$&0.1574 $\pm$ 0.0002&0.1578 $\pm$ 0.0009&\\
20052&2538&photo-non-Ia&$-$&$-$&$-$&0.1574 $\pm$ 0.0002&$-$&\\
20088&2546&Ia&$-$&8.4 $\pm$ 0.2&$-$&0.2444 $\pm$ 0.0005&$-$&zg\\
20097&2587&$-$&$-$&$-$&$-$&0.221 $\pm$ 0.005&$-$&zs\\
20142&2586&Ia&Ia-norm (sn90N)&4.7 $\pm$ 0.4&-13.2&0.3139 $\pm$ 0.0005&
0.32 $\pm$ 0.01&s, zg, f\\
20144&2541&Ia&Ia-norm&1.1 $\pm$ 0.3&-1 $\pm$ 4&0.220 $\pm$ 0.005&
0.225 $\pm$ 0.005&s, zs\\
20227&2568&Ia&Ia-norm&7.5 $\pm$ 0.4&7 $\pm$ 4&0.2764 $\pm$ 0.0005&
0.282 $\pm$ 0.006&s, zg\\
20345&2567&Ia&Ia-norm&-0.7 $\pm$ 0.3&0 $\pm$ 5&0.265 $\pm$ 0.005&
0.267 $\pm$ 0.008&zs\\
20364&2581&Ia&Ia-norm&-1.3 $\pm$ 0.6&0 $\pm$ 5&0.2181 $\pm$ 0.0009&
0.218 $\pm$ 0.007&zg\\
20376&2582&Ia&Ia-norm&3.8 $\pm$ 0.5&2 $\pm$ 6&0.2109 $\pm$ 0.0005&
0.204 $\pm$ 0.007&s, zg\\
20388&2611&photo-non-Ia&$-$&$-$&$-$&0.1787 $\pm$ 0.0005&$-$&zg\\
20430&2543&Ia&Ia-norm&1.4 $\pm$ 0.3&0 $\pm$ 5&0.164 $\pm$ 0.005&
0.168 $\pm$ 0.005&zs\\
20474&2563&$-$&$-$&$-$&$-$&0.2713 $\pm$ 0.0005&$-$&zg, f\\
20474&2714&$-$&$-$&$-$&$-$&0.2713 $\pm$ 0.0005&$-$&zg\\
20474&3003&$-$&$-$&$-$&$-$&0.2713 $\pm$ 0.0005&$-$&zg\\
20530&2547&II&II&$-$&$-$&0.06135 $\pm$ 0.00008&$-$&v\\
20530&2571&II&II&$-$&$-$&0.06135 $\pm$ 0.00008&$-$&v\\
20575&2540&Ia&Ia-norm&2.3 $\pm$ 0.3&0 $\pm$ 4&0.1988 $\pm$ 0.0005&
0.204 $\pm$ 0.005&s, zg\\
20575&3005&Ia&$-$&0.7 $\pm$ 0.3&$-$&0.1988 $\pm$ 0.0005&$-$&s, zg\\
20625&2551&Ia&Ia-norm&-5.4 $\pm$ 0.1&-1 $\pm$ 4&0.1082 $\pm$ 0.0002&
0.108 $\pm$ 0.004&s\\
20625&2604&Ia&Ia-norm&-3.6 $\pm$ 0.1&-1 $\pm$ 4&0.1082 $\pm$ 0.0002&
0.110 $\pm$ 0.004&s\\
20677&2536&Ic&Ib/c&$-$&$-$&0.0804 $\pm$ 0.0001&$-$&v\\
20678&2610&photo-Ia&$-$&3.9 $\pm$ 0.3&$-$&0.2056 $\pm$ 0.0001&$-$&\\
20687&2596&Ia&Ia-norm&-0.2 $\pm$ 0.5&-1 $\pm$ 4&0.1918 $\pm$ 0.0005&
0.194 $\pm$ 0.005&s, zg\\
20687&2597&Ia&Ia-norm&-1.3 $\pm$ 0.5&2 $\pm$ 4&0.1918 $\pm$ 0.0005&
0.191 $\pm$ 0.002&s, zg\\
20718&2577&$-$&Gal&$-$&$-$&0.0888 $\pm$ 0.0001&0.0892 $\pm$ 0.0003&f\\
20718&2593&$-$&$-$&$-$&$-$&0.0888 $\pm$ 0.0001&$-$&\\
20764&2594&Ia&Ia-norm&-2.5 $\pm$ 0.3&-1 $\pm$ 5&0.1664 $\pm$ 0.0005&
0.170 $\pm$ 0.005&s, zg\\
20834&2598&Ia&Ia&-3 $\pm$ 1$^{p}$&$-$&0.1909 $\pm$ 0.0005&$-$&v, zg\\
20862&2600&Ia&Ia-norm&-4 $\pm$ 1$^{p}$&-1 $\pm$ 4&0.2665 $\pm$ 0.0005&
0.266 $\pm$ 0.005&s, zg\\
20909&2580&photo-non-Ia&$-$&$-$&$-$&0.1586 $\pm$ 0.0005&$-$&zg\\
20978&2609&Ia&Ia-pec&-2.6 $\pm$ 0.7$^{p}$&-5 $\pm$ 2&0.324 $\pm$ 0.005&
0.314 $\pm$ 0.001&zs\\
21006&2566&Ia&Ia-norm&1.7 $\pm$ 0.4&0 $\pm$ 4&0.291 $\pm$ 0.005&
0.295 $\pm$ 0.005&zs\\
21033&2565&Ia&Ia-norm&-3.3 $\pm$ 0.4&-2 $\pm$ 4&0.229 $\pm$ 0.005&
0.231 $\pm$ 0.006&zs\\
21034&2719&Ia&Ia&13.3 $\pm$ 0.2&$-$&0.10858 $\pm$ 0.00006&$-$&v\\
21034&2733&Ia&Ia&15.1 $\pm$ 0.2&16 $\pm$ 13&0.10858 $\pm$ 0.00006&
0.111 $\pm$ 0.004&s\\
21042&2564&Ia&Ia-norm&-6.4 $\pm$ 1.0&-1 $\pm$ 5&0.3109 $\pm$ 0.0005&
0.310 $\pm$ 0.007&zg\\
21058&2579&photo-non-Ia&$-$&$-$&$-$&0.1643 $\pm$ 0.0005&$-$&zg\\
21058&2595&photo-non-Ia&$-$&$-$&$-$&0.1643 $\pm$ 0.0005&$-$&v, zg\\
21062&2613&Ia&Ia-norm&-5.3 $\pm$ 0.2&0 $\pm$ 6&0.1388 $\pm$ 0.0001&
0.154 $\pm$ 0.007&s\\
21064&2532&II&II&$-$&$-$&0.07930 $\pm$ 0.00008&$-$&v\\
21064&2533&II&IIP&$-$&1 $\pm$ 3&0.07930 $\pm$ 0.00008&0.077 $\pm$ 0.002&\\
21362&2636&II&$-$&$-$&$-$&0.0867 $\pm$ 0.0005&$-$&zg\\
21362&2697&II&IIP&$-$&8 $\pm$ 27&0.0867 $\pm$ 0.0005&0.082 $\pm$ 0.006&zg\\
21422&2599&Ia&Ia-norm&-4 $\pm$ 1&-2 $\pm$ 5&0.267 $\pm$ 0.005&0.265 $\pm$ 0.008&
zs\\
21502&2574&Ia&Ia-norm&-8.6 $\pm$ 0.2&-1 $\pm$ 4&0.089 $\pm$ 0.001&
0.090 $\pm$ 0.004&s\\
21502&2575&Ia&Ia-norm&-7.7 $\pm$ 0.2&-1 $\pm$ 5&0.089 $\pm$ 0.001&
0.091 $\pm$ 0.004&s\\
21502&2717&Ia&Ia-norm&13.7 $\pm$ 0.2&16 $\pm$ 8&0.089 $\pm$ 0.001&
0.091 $\pm$ 0.004&s\\
21596&2588&photo-non-Ia&$-$&$-$&$-$&0.0633 $\pm$ 0.0002&$-$&f\\
21596&2589&photo-non-Ia&Gal&$-$&$-$&0.0633 $\pm$ 0.0002&0.0673 $\pm$ 0.0008&f\\
21669&2591&Ia&Ia-norm&-8 $\pm$ 2$^{p}$&-3 $\pm$ 4&0.1242 $\pm$ 0.0002&
0.105 $\pm$ 0.006&s\\
21669&2722&Ia&Ia-norm&14 $\pm$ 2$^{p}$&13 $\pm$ 17&0.1242 $\pm$ 0.0002&
0.119 $\pm$ 0.004&s\\
21766&2638&$-$&$-$&$-$&$-$&0.12788 $\pm$ 0.00006&$-$&\\
21810&2724&Ia&Ia&3.0 $\pm$ 0.4$^{p}$&11 $\pm$ 3&0.175 $\pm$ 0.005&
0.180 $\pm$ 0.008&zs\\
21814&2702&Ia&Ia&11.5 $\pm$ 0.2&$-$&0.1021 $\pm$ 0.0005&$-$&v, zg\\
21839&2716&Ia&Ia-norm&4 $\pm$ 1$^{p}$&0 $\pm$ 5&0.0935 $\pm$ 0.0005&
0.096 $\pm$ 0.005&s, zg\\
21861&2723&Ia&Ia-norm&1 $\pm$ 2$^{p}$&-1 $\pm$ 5&0.188 $\pm$ 0.005&
0.192 $\pm$ 0.005&s, zs\\
21898&2704&$-$&$-$&$-$&$-$&0.0388 $\pm$ 0.0002&$-$&\\
22182&2690&Ia&Ia&9 $\pm$ 2$^{p}$&$-$&0.076 $\pm$ 0.005&$-$&v, zs\\
22284&2735&Ia&Ia-norm&1 $\pm$ 3$^{p}$&-1 $\pm$ 5&0.1375 $\pm$ 0.0006&
0.137 $\pm$ 0.006&s, zg\\
\end{longtable}
\tablefoot{
\tablefoottext{a}{The overall SDSS type which is based on the results from the NTT/NOT analysis in combination with spectra from other telescopes.}
\tablefoottext{b}{The typing of the individual spectra in this analysis based on the result from SNID in combination with a visual inspection. If a subtype could be determined, this is given in the table, otherwise only the type. Thus, when the type is given as solely \emph{Ia}, this means that SNID could not determine if it is a normal SN Ia or a non-normal (peculiar, SN 1991T or SN 1991bg). In the cases where all template spectra which could be matched with the input spectrum with SNID belong to the same SN, the name of the SN is written within parenthesis. When the type is \emph{NotSN} or \emph{Gal}, this usually means that the transient was too faint, or the observing conditions were not good enough, so that the contribution from the host galaxy dominates the spectrum.}
\tablefoottext{c}{Number of days in rest frame from $B$-band maximum obtained from the lightcurve. An age is given when the object has been classified as a SN~Ia, as well as there were enough photometry to build a lightcurve. The sign '$p$' is added after the lightcurve age when the lightcurve lacked photometry either before or after maximum brightness.}
\tablefoottext{d}{Number of days in rest frame from $B$-band maximum obtained from SNID. If no error is given, there was only one template fitting the requirements in SNID.}
\tablefoottext{e}{The object redshifts from Zheng et al. (in preparation), which have their origin in SDSS DR7 redshifts in combination with measurements of both NTT/NOT spectra and other spectra.}
\tablefoottext{f}{Additional information. When the SNID type was determined from the host-galaxy subtracted spectrum, this is
marked by an $s$ in the column. If there were few template spectra matching the input spectrum, at which fewer than 5 spectra were used to determine the type, the redshift and/or the age, this is marked with an $f$. When the type obtained from SNID was changed after the visual inspection this is marked with a $v$. When the redshift is not obtained from SDSS DR7, it is marked with \emph{zg} or \emph{zs}, depending on if it was determined from galaxy lines or SN features. If the redshift was determined through template fitting, this is marked with \emph{zt}.}
}
}

\longtab{3}{
\begin{longtable}{cccc}
\caption{\label{tab:qual} Spectral quality.} \\
\hline \hline
ID & SPID & Host Contamination \tablefootmark{a} & Slit Loss \tablefootmark{b} \\
\hline
\endfirsthead
\caption{continued.}\\
\hline\hline
ID & SPID & Host Contamination \tablefootmark{a} & Slit Loss \tablefootmark{b}\\
\hline
\endhead
\hline
\multicolumn{4}{l}{{Continued on next page\ldots}} \\
\endfoot
\hline
\endlastfoot
12778&692&1.0&0.4\\
12779&693&0.4&0.4\\
12781&680&0.1&0.2\\
12782&681&0.9&0.3\\
12820&711&0.6&0.2\\
12842&682&0.1&0.2\\
12843&727&0.3&0.4\\
12844&684&1.0&0.4\\
12853&685&0.1&0.3\\
12855&716&0.7&0.3\\
12856&695&0.5&0.5\\
12860&688&0.2&0.2\\
12874&689&0.2&0.2\\
12898&712&0.5&0.1\\
12907&714&0.8&0.3\\
12927&690&0.5&0.3\\
12928&686&0.3&0.3\\
12930&687&0.2&0.1\\
12947&691&0.7&0.5\\
12950&700&0.3&0.1\\
12950&1055&0.8&0.7\\
12978&701&0.1&0.5\\
13005&702&0.5&0.2\\
13025&761&0.2&0.8\\
13044&724&0.1&0.4\\
13044&1062&0.2&0.8\\
13045&734&0.8&0.6\\
13046&726&1.0&0.8\\
13070&736&0.4&0.1\\
13072&723&0.2&0.2\\
13135&739&0.1&0.4\\
13135&998&0.1&0.3\\
13174&766&0.8&0.1\\
13195&764&0.3&0.7\\
13195&983&0.6&0.5\\
13195&1458&0.6&0.9\\
13355&1003&0.7&0.5\\
13376&1002&0.5&0.5\\
13376&1106&0.5&0.7\\
13577&1000&0.4&0.5\\
13796&1058&0.0&0.1\\
13894&1039&0.3&0.5\\
14157&1040&0.3&0.7\\
14279&1459&0.9&0.7\\
14318&1594&0.4$^{e}$&0.1\\
14318&1653&0.4$^{e}$&0.1\\
14318&1713&0.6$^{e}$&0.3\\
14437&1061&0.4&0.1\\
14450&991&0.5&0.3\\
14451&989&0.6&0.5\\
14492&1001&0.5&0.2\\
14598&987&1.0&0.4\\
14599&988&0.0&0.1\\
14782&990&0.7&0.2\\
14846&1014&0.3&0.2\\
14871&1008&0.4&0.2\\
14979&1009&0.2&0.1\\
14984&1027&0.5&0.2\\
15031&985&0.1&0.3\\
15129&1015&0.3&0.7\\
15132&1012&0.1&0.1\\
15136&1022&0.5&0.1\\
15153&1046&0.1&0.1\\
15161&1010&0.5&0.5\\
15171&1045&0.1&0.1\\
15203&1026&0.6&0.4\\
15207&1038&0.0&0.4\\
15210&1005&0.1&0.3\\
15210&1052&0.1&0.2\\
15222&1004&0.2&0.8\\
15234&1043&0.7&0.6\\
15259&1051&0.2&0.3\\
15287&1057&0.0&0.5\\
15320&1098&0.1&0.1\\
15339&1107&0.9&0.4\\
15354&1110&0.2&0.5\\
15475&1464&1.0$^{e}$&1.0\\
15557&1532&0.0&0.2\\
16021&1355&0.5&0.1\\
16069&1358&0.6&0.8\\
16069&1467&0.5&0.4\\
16069&1651&0.9&0.6\\
16087&1455&0.3&1.0\\
16163&1678&1.0$^{e}$&0.1\\
16165&1326&0.1&0.3\\
16179&1323&0.4&0.6\\
16179&1469&0.2&0.9\\
16179&1570&0.2&1.0\\
16192&1322&0.6&0.5\\
16192&1496&0.7$^{e}$&0.5\\
16204&1500&0.9&0.9\\
16206&1501&0.8&0.7\\
16215&1456&0.4&0.9\\
16215&1630&0.9&0.7\\
16241&1470&0.6$^{e}$&0.9\\
16280&1471&0.7&0.8\\
16280&1564&1.0&0.7\\
16287&1449&0.0&1.0\\
16287&1569&0.0&1.0\\
16287&1650&0.6&0.4\\
16302&1473&0.3&0.4\\
16314&1335&0.1&0.2\\
16314&1475&0.0&0.2\\
16333&1367&0.8&0.4\\
16352&1478&0.1&0.5\\
16391&1452&0.1&1.0\\
16391&1565&0.0$^{e}$&0.6\\
16392&1365&0.3&0.4\\
16392&1448&0.1&0.9\\
16392&1566&0.5$^{e}$&0.5\\
16392&1682&0.4$^{e}$&0.2\\
16402&1505&0.4&1.0\\
16473&1520&0.3&0.1\\
16541&1485&0.2&0.4\\
16578&1516&0.6$^{e}$&0.8\\
16619&1519&0.0&0.6\\
16619&1528&0.1&0.2\\
16637&1514&0.5&1.0\\
16641&1518&0.2&0.9\\
16641&1530&0.4&0.3\\
16641&1649&0.4$^{e}$&0.8\\
16668&1561&0.1&0.2\\
16692&1489&0.1&0.1\\
16737&1599&0.1$^{e}$&1.0\\
16741&1523&0.0&0.2\\
16748&1574&0.4$^{e}$&0.2\\
16774&1606&0.1&0.7\\
16778&1542&0.4&0.0\\
16778&1568&0.1&0.1\\
16793&1603&0.0&0.9\\
16838&1522&0.0&0.4\\
16857&1538&0.7&1.0\\
16867&1541&0.2&0.6\\
16872&1539&0.0$^{e}$&0.8\\
16956&1562&0.2&0.1\\
16979&1597&0.2&0.8\\
16988&1595&0.8&1.0\\
16988&1652&0.8$^{e}$&1.0\\
17117&1679&0.0$^{e}$&0.3\\
17135&1648&$-$&0.4\\
17167&2250&0.6&0.5\\
17170&1879&0.8&0.4\\
17176&1812&1.0&0.3\\
17200&1796&0.9&0.5\\
17206&1788&0.6&0.4\\
17218&1794&0.8&0.3\\
17220&1791&0.9&0.1\\
17223&1793&0.3&0.2\\
17237&1830&0.6&0.2\\
17245&2234&0.5&0.6\\
17247&1799&0.7&0.2\\
17253&1898&0.7&0.4\\
17254&1813&0.9&0.3\\
17332&1899&0.0&0.4\\
17351&1769&0.9&0.4\\
17366&1782&0.0&0.2\\
17389&1811&0.6&0.3\\
17391&1872&0.1&0.1\\
17422&1785&0.1&0.3\\
17435&1902&0.1&0.4\\
17436&1790&0.4&0.3\\
17464&1853&0.5&0.3\\
17486&1854&1.0&0.3\\
17497&1837&0.4&0.3\\
17500&2249&0.0&0.4\\
17535&1838&0.4&0.3\\
17548&1825&0.7&0.1\\
17548&2231&0.6&0.5\\
17548&2293&0.8&0.2\\
17552&1789&0.3&0.1\\
17568&1810&0.6&0.1\\
17605&1809&1.0&0.2\\
17627&1841&0.7&0.3\\
17629&1851&0.8&0.4\\
17647&1875&0.2&0.7\\
17703&1881&0.2&0.2\\
17745&2161&0.6&0.1\\
17746&1873&0.1&0.2\\
17784&1842&0.0&0.5\\
17790&1887&0.0&0.6\\
17794&1906&0.0&0.4\\
17811&1816&0.0&0.2\\
17814&1901&0.8&0.5\\
17825&1819&0.1&0.2\\
17854&2230&0.0&0.7\\
17875&1817&0.4&0.3\\
17880&1843&0.1&0.4\\
17880&1957&0.4&0.2\\
17880&2194&0.8&0.6\\
17886&1844&0.1&0.3\\
17924&1826&1.0&0.3\\
17973&1926&0.9&0.3\\
17973&1942&1.0&0.1\\
18109&1940&0.3&0.3\\
18325&2277&0.2&0.5\\
18457&2285&0.8&0.2\\
18466&2270&0.1&0.2\\
18590&2248&0.9&0.7\\
18596&2227&0.1&0.3\\
18647&2271&0.5&0.2\\
18697&2171&0.7&0.2\\
18768&2135&0.1&0.5\\
18787&2150&0.2&0.5\\
18804&2148&0.2&0.5\\
18903&2247&0.8&0.5\\
18965&2279&0.6&0.5\\
19003&2235&0.9&0.9\\
19003&2290&0.9&0.5\\
19008&2284&0.8&0.2\\
19023&2236&0.0&0.4\\
19051&2297&0.5&0.2\\
19101&2268&0.1&0.2\\
19149&2275&0.0&0.4\\
19155&2252&0.3&0.9\\
19155&2607&0.3&0.0\\
19155&2720&0.8&0.8\\
19221&2274&0.5&0.5\\
19222&2299&0.2&0.2\\
19230&2282&0.9&0.4\\
19282&2280&0.1&0.4\\
19323&2296&0.5&0.6\\
19341&2298&0.3&0.1\\
19353&2281&0.3&0.4\\
19381&2283&0.1&0.5\\
19899&2550&0.0&0.0\\
19913&2585&0.4&0.1\\
19953&2602&0.2&0.0\\
19968&2549&0.1&0.1\\
20039&2584&0.6&0.7\\
20040&2612&0.4&0.8\\
20052&2537&0.1&0.0\\
20052&2538&0.2&0.9\\
20088&2546&0.8&0.7\\
20097&2587&0.1&0.5\\
20142&2586&0.4&0.7\\
20144&2541&0.2&1.0\\
20227&2568&0.7&0.6\\
20345&2567&0.1&1.0\\
20364&2581&0.4&1.0\\
20376&2582&0.8&0.1\\
20388&2611&0.6&0.8\\
20430&2543&0.1&1.0\\
20474&2563&0.4&0.5\\
20474&2714&0.6$^{e}$&0.1\\
20474&3003&0.7$^{e}$&0.1\\
20530&2547&0.9&0.6\\
20530&2571&0.8&0.8\\
20575&2540&0.9&0.1\\
20575&3005&0.8&0.1\\
20625&2551&0.4&0.0\\
20625&2604&0.3&0.0\\
20677&2536&0.9&0.7\\
20678&2610&0.8&0.1\\
20687&2596&0.9&0.1\\
20687&2597&0.9&0.1\\
20718&2577&0.9&0.4\\
20718&2593&0.8&0.1\\
20764&2594&0.7&0.1\\
20834&2598&0.9&1.0\\
20862&2600&0.7&0.1\\
20909&2580&0.8&1.0\\
20978&2609&0.2&1.0\\
21006&2566&0.0&0.8\\
21033&2565&0.1&0.7\\
21034&2719&0.4&0.8\\
21034&2733&0.7&0.3\\
21042&2564&0.4&0.8\\
21058&2579&0.5&0.8\\
21058&2595&0.5&0.1\\
21062&2613&0.3&0.9\\
21064&2532&0.2&0.1\\
21064&2533&0.3&0.5\\
21362&2636&0.3&0.0\\
21362&2697&0.3$^{e}$&1.0\\
21422&2599&0.1&0.3\\
21502&2574&0.3&0.0\\
21502&2575&0.5&0.1\\
21502&2717&0.7&0.4\\
21596&2588&0.8&0.1\\
21596&2589&0.8&1.0\\
21669&2591&0.2&0.7\\
21669&2722&0.1$^{e}$&0.3\\
21766&2638&1.0&0.0\\
21810&2724&0.1$^{e}$&0.1\\
21814&2702&0.9&0.6\\
21839&2716&0.4$^{e}$&0.1\\
21861&2723&0.2$^{e}$&0.9\\
21898&2704&1.0$^{e}$&0.9\\
22182&2690&0.0$^{e}$&0.7\\
22284&2735&0.6$^{e}$&0.3\\
\end{longtable}
\tablefoot{
 Both these quantities are very rough estimates and are in most cases likely overestimated. \\
\tablefoottext{a}{The estimated host-galaxy contamination in the observed spectra, i.e. before host-galaxy subtraction has been attempted, for wavelengths corresponding to the {\contfilter} filter in observed frame. A superscript of '$e$' is used when the estimation is based on an extrapolation of the photometry.}
\tablefoottext{b}{The estimated maximum differential slit loss due to atmospheric refraction affecting the SN light in the wavelength region 4000-8000 {\AA} (observed frame).}
}
}

\end{document}